\documentclass[twocolumn]{aastex631}  %twocolumn
\usepackage{enumitem}
\usepackage{xcolor}
\usepackage{tablefootnote}
\usepackage{bm}
\usepackage{lineno}
\hypersetup{backref=true,  pagebackref=true,  hyperindex=true, colorlinks=blue,	breaklinks=true,  urlcolor=violet,  linkcolor= red, bookmarks=true, 	bookmarksopen=true,  filecolor=cyan, citecolor=blue, linkbordercolor=blue}

\newcommand\wfirst{\textit{Roman}}

\shorttitle{Wavelet transforms of microlensing data}
\shortauthors{Sajadian, et. al.,}

\begin{document}
\title{Wavelet transforms of microlensing data: Denoising, extracting intrinsic pulsations, and planetary signals}
\author[0000-0002-0167-3595]{Sedighe~Sajadian$^{a}$}
\affiliation{Department of Physics, Isfahan University of Technology, Isfahan 84156-83111, Iran}

\author[0000-0002-7611-9249]{Hossein Fatheddin}
\affiliation{Leiden Observatory, Leiden University, PO Box 9513, NL-2300 RA Leiden, the Netherlands}

\footnote{$^{a}$  \textcolor{blue}{$\rm{e}$-$\rm{mail:s.sajadian@iut.ac.ir}$}}

\begin{abstract}
Wavelets are waveform functions that describe transient and unstable variations, such as noises. In this work, we study the advantages of discrete and continuous wavelet transforms (DWT and CWT) of microlensing data to denoise them and extract their planetary signals and intrinsic pulsations hidden by noises. We first generate synthetic microlensing data and apply wavelet denoising to them. For these simulated microlensing data with ideally Gaussian nosies based on the OGLE photometric accuracy, denoising with DWT reduces standard deviations of data from real models by $0.044$-$0.048$ mag. The efficiency to regenerate real models and planetary signals with denoised data strongly depends on the observing cadence and decreases from $37\%$ to $0.01\%$ by worsening cadence from $15$ min to $6$ hrs. We then apply denoising on $100$ microlensing events discovered by the OGLE group. On average, wavelet denoising for these data improves standard deviations and $\chi^{2}_{\rm n}$ of data with respect to the best-fitted models by $0.023$ mag, and $1.16$, respectively. The best-performing wavelets (based on either the highest signal-to-noise ratio's peak ($\rm{SNR}_{\rm{max}}$), or the highest Pearson's correlation, or the lowest Root Mean Squared Error (RMSE) for denoised data) are from 'Symlet', and 'Biorthogonal' wavelets families in simulated, and OGLE data, respectively. In some denoised data, intrinsic stellar pulsations or small planetary-like deviations appear which were covered with noises in raw data. However, through DWT denoising rather flattened and wide planetary signals could be reconstructed than sharp signals. CWT and 3D frequency-power-time maps could advise about the existence of sharp signals. 
\end{abstract}


%%%%%%%%%%%%%%%%%%%%%%%%%%%%%%%%%
\section{Introduction}
Gravitational microlensing, i.e., the temporary brightening of a background source star because its light is passing from the gravitational potential of a collinear and massive object \citep{Einstein1936,1964MNRASrefsdal, Liebes1964}, is a method effective for discovering extrasolar planets \citep{1991Maoplanet, Gould1992}. This method is sensitive to exoplanets mostly outside the snowlines of their parent stars, and further than $\sim 1$ kpc from the observer \citep[see, e.g., ][]{2012Gaudireview}. Detecting such exoplanets is only possible through gravitational microlensing observations.    

In planetary microlensing events, usually a planet is rotating the lens object and gravitationally perturbs the light of images. The resulted planetary signals are usually small and short, and their durations are proportional to $\sqrt{q}$, where $q$ is the planet/host star mass ratio \citep[e.g., ][]{Gould1992}. For example, in a common microlensing event caused by an M dwarf lensing object with an orbiting Earth-mass planet toward the Galactic bulge \citep[e.g., ][]{2008dominik}, the time period of its planetary signal is $\sim \sqrt{q}~t_{\rm E}\sim 1.5$ hours. Here, we assume for this common microlensing event the Einstein crossing time $t_{\rm E}=20$ days. 

\noindent The peak of a planetary signal strongly depends on $\rho_{\star}$ (which is the source radius projected onto the lens plane and normalized to the Einstein radius) as well as the caustic size. The Einstein's radius is the radius of the fully aligned image ring. The larger projected source radii, the more flattened planetary signals \citep{1994witt,2023sajadian}.     

To improve detecting these planetary signatures in microlensing events, two factors need to be improved, namely (i) the observing cadence, and (ii) the photometric accuracy of the data. 

\noindent The former can be improved by either follow-up observations with other telescopes or dense observations with survey telescopes such as the {\it Nancy Grace Roman Telescope} (\wfirst)\ telescope which is scheduled to observe the Galactic bulge during six $62$-day seasons in its $5$-year mission \citep{2019Penny}. Nowadays, several follow-up groups monitor ongoing microlensing events alerted by survey telescopes to cover entire planetary signals, including MiNDStep\footnote{\url{www.mindstep-science.org}}, PLANET\footnote{\url{http://planet.iap.fr/}}, RoboNet\footnote{\url{https://robonet.lcogt.net/}}, MicroFUN \footnote{\url{https://cgi.astronomy.osu.edu/microfun/}}\citep{Dominik2010,1999astro.ph.10465D,2009Tsapras}.

\noindent Concerning the second factor, the photometric accuracy of microlensing data depends on the observing device. For instance, the photometric accuracy of {\it The Optical Gravitational Lensing Experiment} (OGLE\footnote{\url{https://ogle.astrouw.edu.pl/}}) \citep{OGLE2003_1,OGLE_IV} microlensing observations is limited to $\sim 0.003$ mag, and the \wfirst\ accuracy in the W149 filter is planned to reach $\sim0.001$ mag \citep{2019Penny}. To improve the photometric accuracy it is helpful using either Lucky Imaging camera, or adaptive optics, or active optics \citep[e.g. see ][]{1993AdaptiveO,1978activeO,2012Harpsoe}. All of these methods are capable to capture rapid atmospheric turbulences with time scales shorter than the image exposure time, and remove them. Another method to reduce noises is the wavelet transform of data. Wavelets can model transient and unstable noises happen during a limited time interval (e.g. overnight).

Wavelets are waveform and localized mathematical functions with finite energy, and zero mean\citep{daubechies1992}. Similar to the Fourier expansion one can expand a given function versus wavelet bases. Since wavelets can model transient and unstable noises, one can denoise time-series data and extract signals using wavelet decomposition \citep{Morlet1982a,Morlet1982b}. Denoising process through DWT includes three steps: (i) decomposition of time-series data into wavelet terms, (ii) applying a given threshold to the wavelet coefficients, and (iii) regenerating the signal using remnant terms. In addition, there are some other powerful applications of wavelets including differential and ordinary equations solving, data compression and signals processing, etc., in various scientific fields such as geophysics, neural engineering, medicine, signal interpretation, astronomy and astrophysics, etc.\citep[see, for example,][]{Murtagh2002,Mallat2006,Radhakrishnan_2018,geophysicswave,Baldazzi_2020,2015Mertens}. 

\noindent Accordingly, wavelet transforms of microlensing data will potentially have positive impacts on (i) detectability of noisy planetary signals, (ii) discerning intrinsic stellar pulsations, and (iii) correctly data modeling by denoising. In this work, we study these points, by applying wavelet transforms on simulated and real microlensing data.  

The plan of the paper is as follows. In Section \ref{form2}, we review the formalism of wavelets, and its applications. In Section \ref{dwt}, we study noise reduction via discrete wavelet transforms of simulated and real microlensing data, and discuss on its significant advantages. In Section \ref{cwt}, we study benefits of the continuous wavelet transform to reveal the existence of sharp planetary signatures. In the last section, \ref{conclusion}, we summarize the results and conclusions.

\section{Wavelet formalism}\label{form2}%% history%%%%%%%%%%%%%%%%%
In this section, we first review historical achievements in the wavelets field, then explain the conventional formalism for the wavelet decomposition, and finally finish this section by inspecting the wavelet's applications.  

The simplest form for a mother wavelet was first introduced by A. Haar (1909), in his doctoral thesis as a set of orthogonal bases. The Haar wavelet has zero mean. Later, P. Levy (1930) improved the Haar wavelets into scale-varying basis functions. 

The first application of wavelets, i.e. decomposing a function into wavelet bases and extracting its net signal from noises, was introduced by \citet{Morlet1982a,Morlet1982b} in a geophysical context, and then followed by \citet{Grossmann1984}. In the regard of wavelet-based decomposition, we know that Joseph Fourier (1807) established a method to represent signals using a series of sine and cosine terms, known as the Fourier transform. Similar to the Fourier transform, we can expand a function to an orthonormal wavelet sequence generated by a mother wavelet. While the Fourier transform analyzes a signal in the 2D frequency-power domain, the wavelet transform takes a signal into the 3D frequency-power-time domain. \citet{Morlet1982a,Morlet1982b} used the windowed Fourier terms (sine and cosine functions with a Gaussian window) as the wavelet bases. 

\noindent Y. Mayer and S. Mallat introduced another new concept in discrete wavelet transform (DWT), which is multi-resolution analysis (MRA) \citep{Mallat1989,meyer_1993,Mallat2006}. In their formalism, there is a scaling function that generates different wavelet bases depending on a certain criterion. A famous family of orthonormal wavelet bases, Daubechies, was introduced by \citet{daubechies1992} using the concept of DWT.

%%%%%%%%% defenision  %%%%%%%%%%%% 
Wavelets are localized mathematical functions with zero mean and finite energy. They have non-zero values for a limited time interval, and zero at other times\citep[see, e.g., ][]{daubechies1992}. Hence, they can model transient and unstable nosies in a better way than the Fourier transform\citep[e.g., ][]{Radhakrishnan_2018}. 

\noindent There are some defined mother wavelets, e.g. Haar, Daubechies, Meyer, Biorthogonal, etc. From each mother wavelet a number of child wavelets can be generated using dilation and translation. The dilation and translation parameters change either continuously or discretely, leading to continuous and discrete wavelet transforms. In the wavelet transform, a given function is expanded versus child wavelets, like: 
\begin{eqnarray}
f(t)= \sum_{a, b} C_{a, b} \psi_{a, b}(t),
\label{expand} 
\end{eqnarray} 
where $C_{a, b}$s are the expansion coefficients, $\psi_{a, b}$s are the child wavelets originated from the mother wavelet $\psi$, by varying its scale (the dilation) and transferring (translation). A general form for a child wavelet basis originated from a mother wavelet $\psi(t)$ is:  
\begin{eqnarray}
\psi_{a, b}= \frac{1}{\sqrt{|a|}}\psi\big(\frac{t-b}{a}\big), 
\end{eqnarray}
where $a$, and $b$ are the so-called dilation and translation coefficients, respectively. Note that each mother wavelet $\psi(t)$ must satisfy the following conditions:
\begin{eqnarray}
\int_{-\infty}^{\infty} \big|\psi(t)\big|^{2} dt&<& \infty, \nonumber\\
\int_{-\infty}^{\infty} \psi(t) dt&=& 0,
\end{eqnarray} 
which ensure two conditions of finite energy and zero average, respectively.

\noindent In the continuous wavelet transform (CWT), the expansion coefficients (in Equation  \ref{expand}) are determined by: 
\begin{eqnarray}
C_{a, b}=\frac{1}{\sqrt{|a|}} \int_{-\infty}^{\infty} f(t)~\psi^{\ast}\big(\frac{t-b}{a}\big) dt. 
\end{eqnarray} 
In DWT, the possible values of $a$, and $b$ are discrete and vary on a dyadic grid like $a= 2^{\pm i}$, and $b=k~a$, where $i$ defines the decomposition level. By doubling the scale parameter $a$, the frequency of the wavelet basis decreases by $1/2$. Depending on a desired decomposition level, we get an \textit{approximated} part for the signal. On the other hand, by reducing $a$ we obtain a \textit{detailed} part of the signal. In multi-resolution analysis, a signal composites of both parts. Therefore, data analysis with DWT can be done in two regimes: low and high frequencies. High-frequency DWT distinguishes sharp and local patterns, and their locations in time-series data, while low-frequency DWT models large-scale signals. 
\begin{figure*}
	\centering	
	\includegraphics[width=0.49\textwidth]{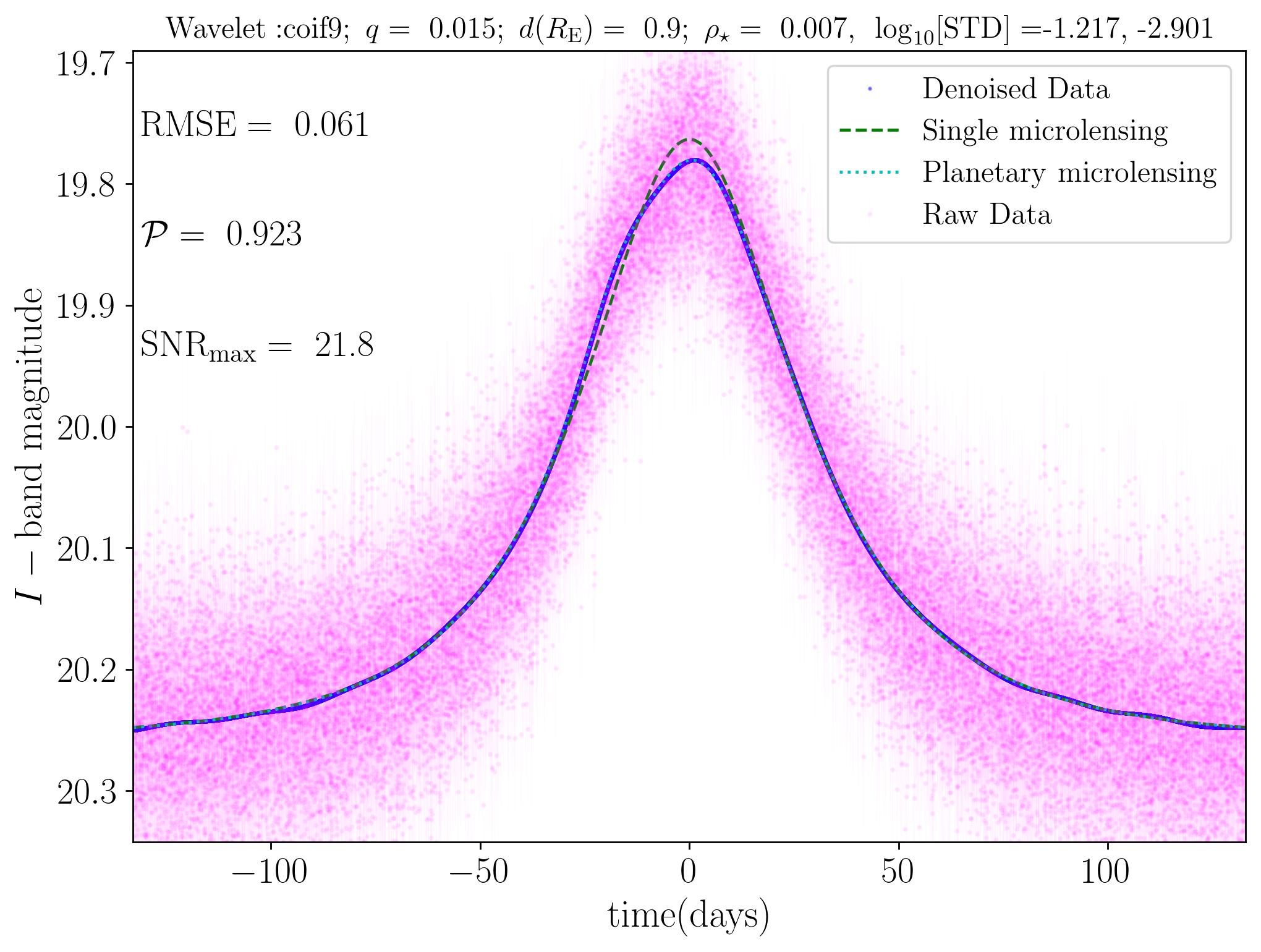}
	\includegraphics[width=0.49\textwidth]{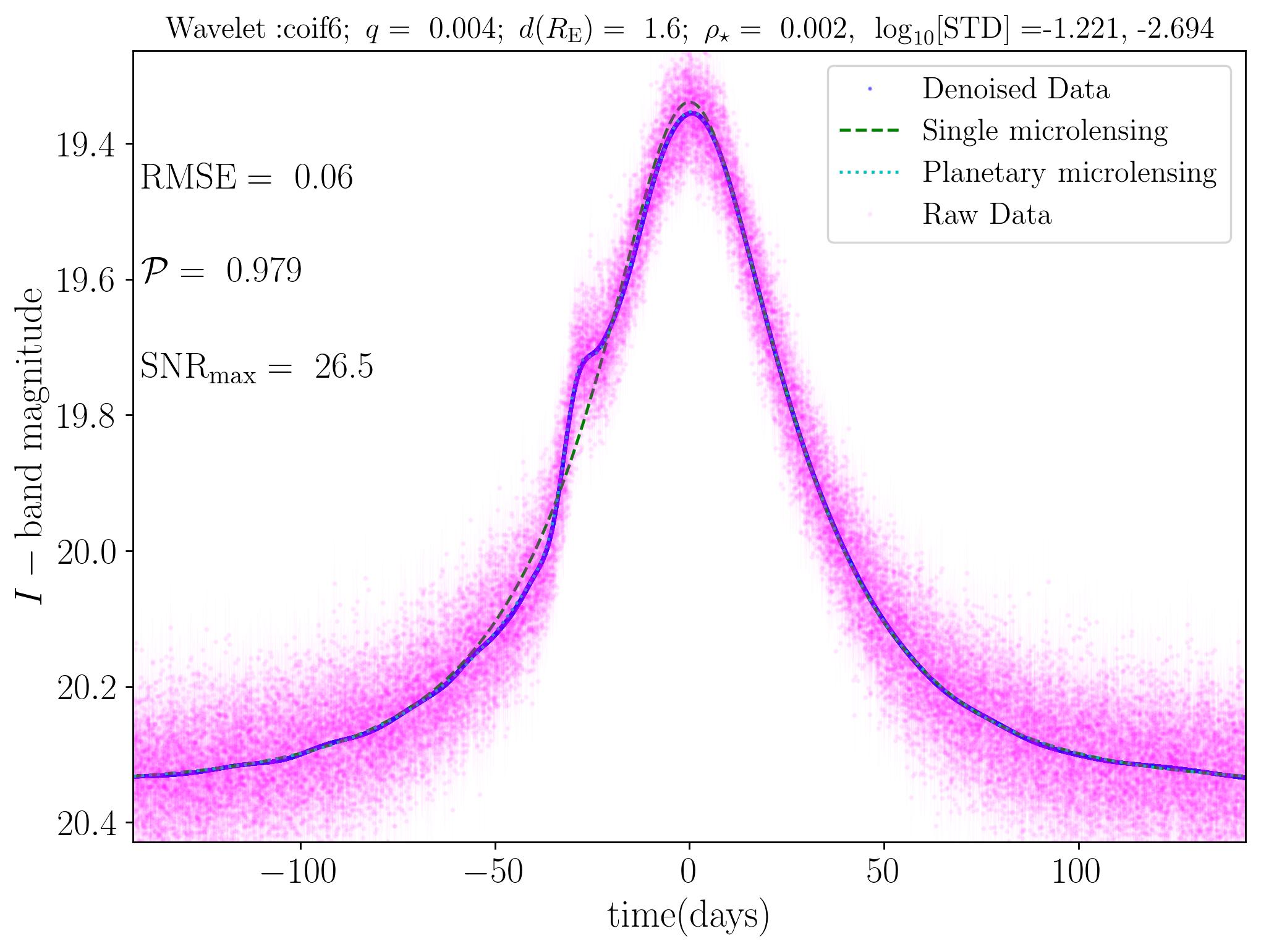}
	\caption{Two examples of simulated planetary microlensing light curves (dotted cyan curves), and the corresponding single microlensing events (by putting $q=0$, dashed green curves). In each panel, the raw, and denoised data are represented with pink and blue colors, respectively. The best-performing wavelet for denoising and the used planetary lensing parameters are mentioned at the top of panels. Two given amounts for $\log_{10}[\rm{STD}\rm{(mag)}]$ are corresponding to raw and denoised data, respectively.}\label{Figone}
\end{figure*}

%% applications
Wavelet analysis has several significant applications, which are solving ordinary and partial differential equations \citep{swe:phd,CRUZ20013305}, data compression, signal analysis (including denoising, compression, signal detectuion), etc. 

\noindent Data compression is based on decomposing a given function into wavelets, and then reducing the correlations in data. 

\noindent To denoise data via DWT, two parameters should be specified namely threshold, and thresholding method. A threshold must be chosen based on the noise information \citep[see example,][]{91217compression,952803compression}. A given threshold is compared to the standard deviation (STD) due to each decomposition level. Decomposition levels with STDs lower than that given threshold are removed from the data. There are two methods for removing these wavelet terms, hard and soft thresholding methods. In the hard method, these wavelets are ignored in the decomposition of data, and in soft method, the given threshold is diminished from all other wavelets. The first makes discontinuity in data and the second degrades the signal, although it does not produce any discontinuity \citep{thresholding}. Hence, the denoising process consists of three steps: decomposition of an input data in wavelet terms, removing wavelets with an STD less than that given threshold, reconstruct the signal using the remnants. Here, in this work we use the denoising application of the wavelet transform for microlensing data.  

\noindent In following subsection, we review some wavelet applications perused in different astronomical/astrophysical fields.    

\subsection{Wavelet applications in Astronomy/Astrophysics}
%Wavelet are non-stationary and localized function, and hence model transient astrophysical events and unstable noises well. 
Up to now, wavelets were used in different astronomical/astrophysical research area, e.g., gamma-ray bursts, blazars, quasars, AGN, pulsating stars, and more \footnote{A comprehensive review on 'Wavelets in the astronomical context' was developed by Sh. Ganesh which is:  \url{https://www.prl.res.in/~shashi/astro_wavelets/report.pdf}} \citep[e.g., ][]{2003GRBwavelet, 2008Blazarwavelet,2023AGNwavelet}. In most of these fields, a phenomenon is described using time-series data, and a DWT process separates signals and noises (which are transient and unstable). 

Additionally, wavelets can improve the signal-to-noise ratio (SNR) for astronomical images \citep[e.g.,  ][]{2015Mertens}. Morphological features in astronomical images are extracted using wavelet multi-scale vision model \citep{Murtagh2002}. In this model, denoising thresholds at different scales are determined independently. 
Analysis of the cosmic microwave background (CMB) maps (including point source detection, map compression, and power spectrum estimation) using wavelets was performed in \citet{cmbwavelet}. 
In addition, analyzing light curves of \textit{Kepler} and CoRoT stars as well as distinguishing rotation, magnetic activity, and pulsating signals were accomplished using CWT and by making 3D frequency-power-time maps in \citet{2014aabravo}. 

Considering different wavelet applications in astronomy and astrophysics, in the next section we study advantages of DWT, and denoising of microlensing data.

\section{DWT of simulated and real microlensing data}\label{dwt}
In this section, we study the advantages of denoising (a) simulated microlensing data (in Subsection \ref{simul}), and (b) real microlensing data (in Subsection \ref{real}). For synthetic microlensing data with ideally Gaussian noises we know their real models. Accordingly, we (i) study the usefulness of denoising process for gravitational microlensing data, and (ii) find the best method for denoising similar data without having real models.

\subsection{Simulated microlensing data}\label{simul}
To generate synthetic planetary and single microlensing light curves, these parameters should be specified: the lens impact parameter $u_{0}$, the time of the closest approach $t_{0}$, the Einstein crossing time $t_{\rm E}$, the normalized source radius projected onto the lens plane $\rho_{\star}$, the baseline apparent magnitude at the source position $m_{\rm{base}}$, the blending parameter $f_{\rm b}$, the planet/host star mass ratio $q$, the planet/host star distance projected on the sky plane and normalized to the Einstein radius $d(R_{\rm E})$, and the angle between the source trajectory and the binary axis $\xi$. 
To produce all possible planetary light curves, we choose these parameters uniformly as $u_{0}\in [0,~1]$, $t_{\rm E} \in [20,~60]$ days, $\log_{10}[\rho_{\star}]\in [-3,~-1.5]$, $\log_{10}[q] \in [-3.5,~-1.5]$, $d(R_{\rm E}) \in [0.9,~1.7]$, $m_{\rm{base}} \in [18.5,~21]$mag, $f_{\rm b} \in [0,~1]$, and $\xi\in[0, 360^{\circ}]$. We set $t_{0}=0$ for all simulated events, and determine the photometric errors according to the OGLE observation \citep[see, Fig. (1) of][]{2021sajadian}. We generate data points by considering different cadences, and assuming continuous observations during $[-3.5~t_{\rm E},~3.5~t_{\rm E}]$. 
We generate 1000 events so that this ensemble includes different lensing configurations and we can visually investigate every denoised event. Simulating planetary microlensing light curves are done using RT-model \citep{Bozza2010,Bozza2018}.

For denoising each simulated event, we use the python module \texttt{scikit}-\texttt{image}\footnote{\url{https://github.com/scikit-image}} \citep{Donohobiometrika,862633Chang}. For each light curve, we apply different wavelets from wavelet families 'Haar(haar)', 'Daubechies(db)', 'Symlets(sym)', 'Coiflets(coif)', 'Biorthogonal(bior)', 'Reverse biorthogonal(rbio)',  'FIR approximation of Mayer wavelet(dmey)',  'Gaussian wavelets(gaus)',  'Complex Gaussian wavelets (cgau)',  'Shannon wavelets(shan)', 'Frequency B-Spline wavelets(fbsp)',  'Mexican hat wavelet (mexh)', and 'Complex Morlet wavelets (cmor)'. These wavelet families are sorted in the python module \texttt{PyWavelets} \footnote{\url{https://github.com/PyWavelets}} \citep{Pywavelets}. 

\noindent The default value for the wavelet levels for denoising is three less than the maximum number of the decomposition level for each wavelet (which itself depends on the number of data points). 

\noindent Possible thresholding methods are 'BayesShrink' (BS, implement different thresholds while wavelet soft thresholding), 'VisuShrink' (VS, consider a single and universal threshold for all wavelet coefficients). For VS methods it is better to estimate a single amount for noises. Suppressing a universal noise in the VS method can be done in soft mode (adds additional noise to the data, but will offer the best approximation of the signal), or hard mode (removing noise which increases SNR, but makes discontinuity in the data).  
\begin{deluxetable}{ccccc}
	\tablecolumns{5}
	\centering
	\tablewidth{0.49\textwidth}\tabletypesize\footnotesize
	\tablecaption{The results from simulated microlensing events denoised with DWT.}\label{tab1}
	\tablehead{\colhead{Cadence}&\colhead{$\epsilon[\%]$}&\colhead{$\overline{\Delta \rm{STD}}(\rm{mag})$}& \colhead{Wavelet}&\colhead{Thresholding}}
	\startdata	
	$15~\rm{min}$&36.8&0.048 & sym7 & VS, soft \\ 
	$1~\rm{hr}$     & 8.7 & 0.047& sym17 & BS, soft \\
	$2~\rm{hrs}$     & 1.9&0.047& sym7 &  BS, soft\\
	$4~\rm{hrs}$      &0.2 & 0.045 & sym7 & BS, soft\\
	$6~\rm{hrs}$      &0.01 & 0.044 & db4 & BS, hard\\
	\enddata
	\tablecomments{In the second column, $\epsilon$ is the fraction of events with $\log_{10}[\rm{STD}(\rm{mag})]<-2.79$ to the total number of simulated events. The next column gives $\overline{\Delta \rm{STD}}(\rm{mag})$ which is the average improvement in data STD (given by Eq. \ref{std}) due to denoising. Two last columns resresenet the most-frequent wavelets and thresholding methods for the best denoised data sets.}
\end{deluxetable}

To find the best wavelet and the best thresholding method while denoising every simulated microlensing data we have three criteria. The first one is the Root Mean Squared Error (RMSE), which is calculated as follows:  
\begin{eqnarray}
\rm{RMSE}= \sqrt{\frac{1}{N}\sum_{i=1}^{N} \big(y_{i}-y_{\rm d,~i}\big)^{2}},
\label{cri1}
\end{eqnarray} 
where, $y_{\rm d,~i}$ is the source apparent magnitude at time $t_{i}$ after denoising, and $y_{i}$ is the original source apparent magnitude on that time. $N$ is the number of data points. A lower RMSE value means that denoised data set is more similar to the original data.  

The second one is the Pearson's correlation \citep{Pearson1895}, which is given by: 

\begin{eqnarray}
\mathcal{P}=\frac{\sum\limits_{i=1}^{N} (y_{i}-\overline{y}) {\bf \textbf{.}}(y_{\rm{d},~i}-\overline{y_{\rm d}})}{\sqrt{ \sum\limits^{N}_{j=1}\big(y_{j}-\overline{y}\big)^{2} {\bf \textbf{.}}\sum\limits_{k=1}^{N} \big(y_{\rm{d},~k}-\overline{y_{\rm d}}\big)^{2}}},
\label{cri2}
\end{eqnarray}
where, $\overline{y}$ and $\overline{y_{\rm d}}$ are the average values of $y$s and $y_{\rm d}$s. $\mathcal{P}$ can change in the range $[-1,~+1]$. A larger value of the Pearson's correlation means that the denoised data set has a more correlation with the original data. 
\begin{figure*}
\centering
\includegraphics[width=0.49\textwidth]{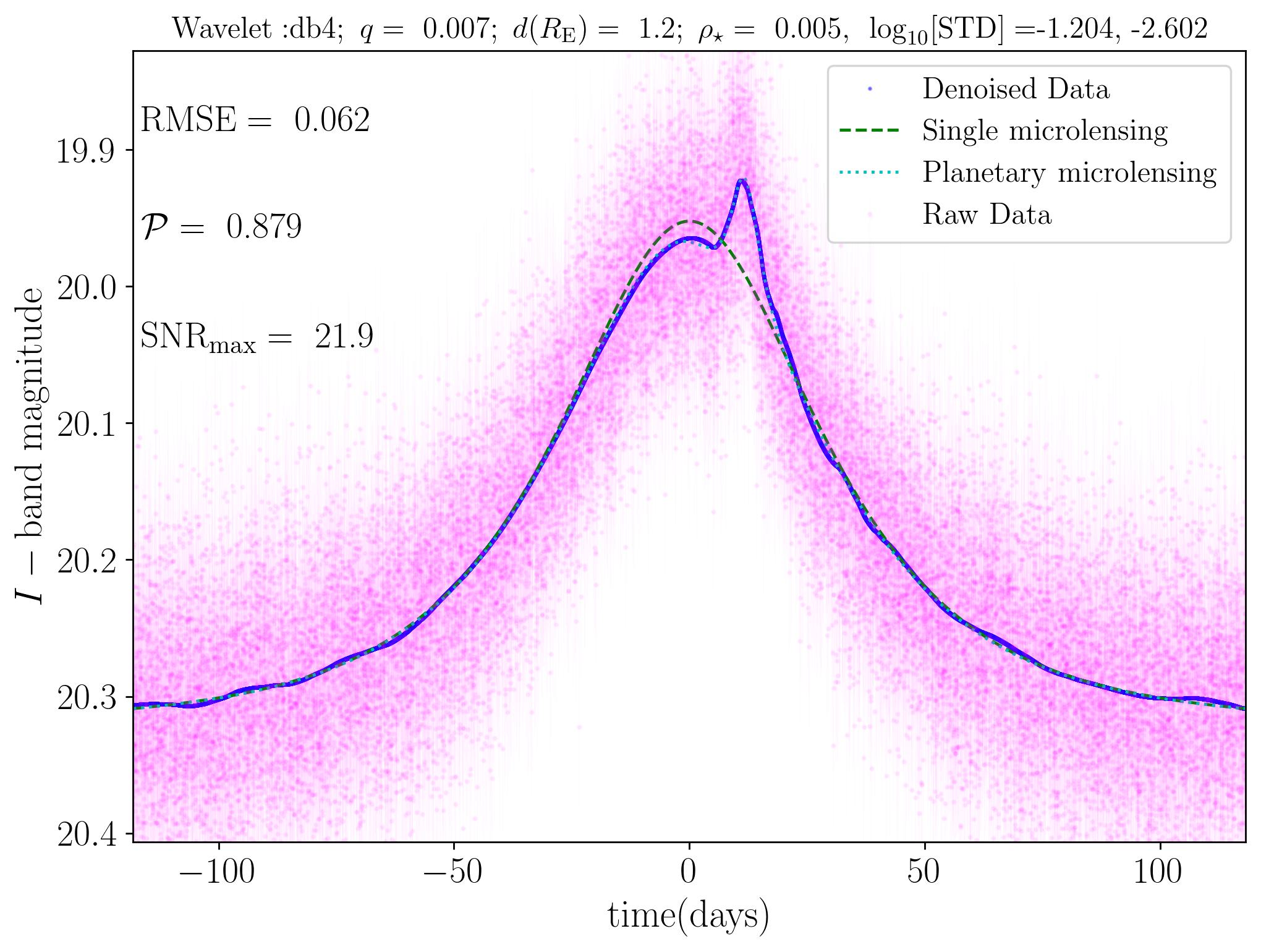}
\includegraphics[width=0.49\textwidth]{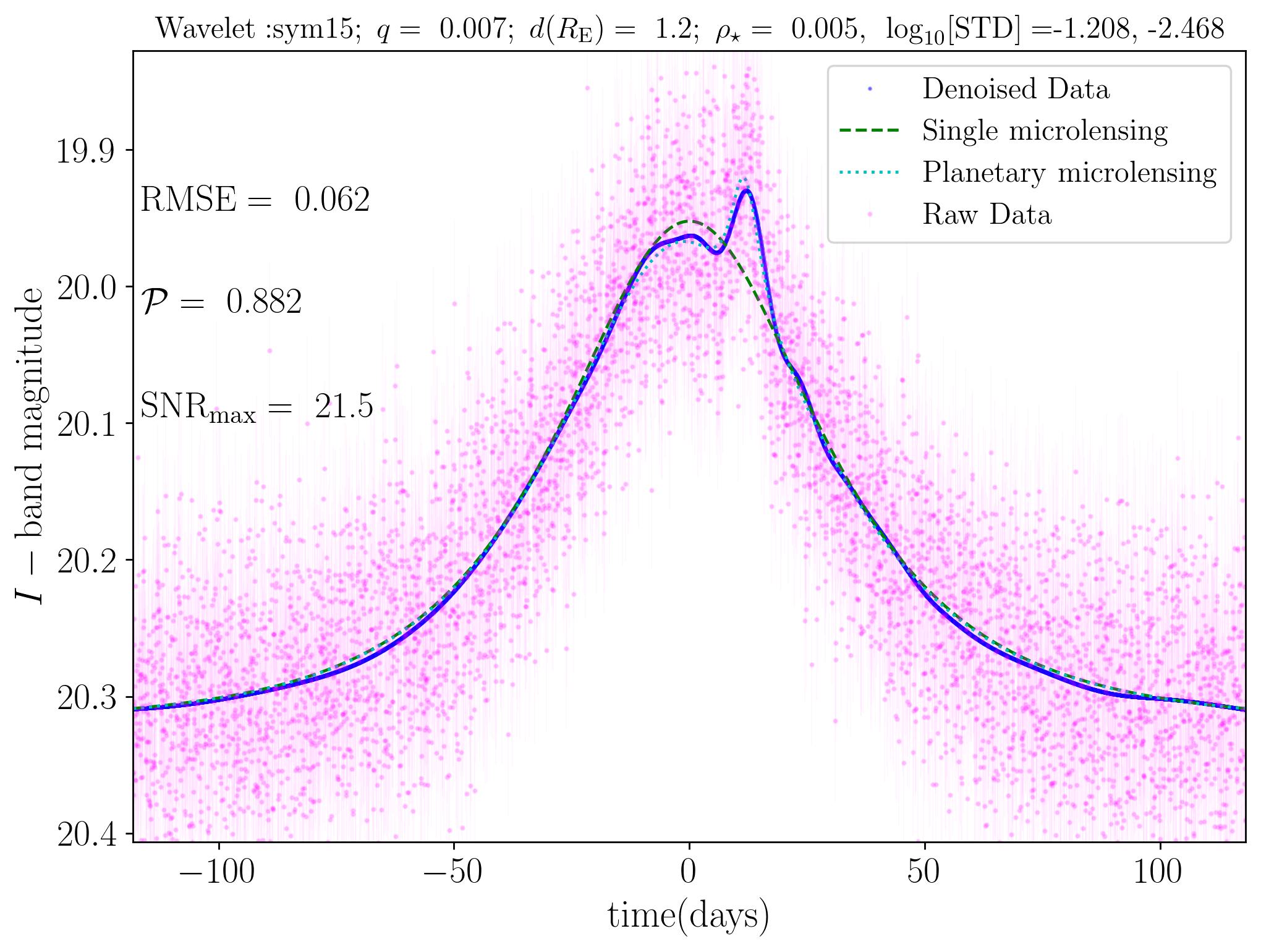}
\includegraphics[width=0.49\textwidth]{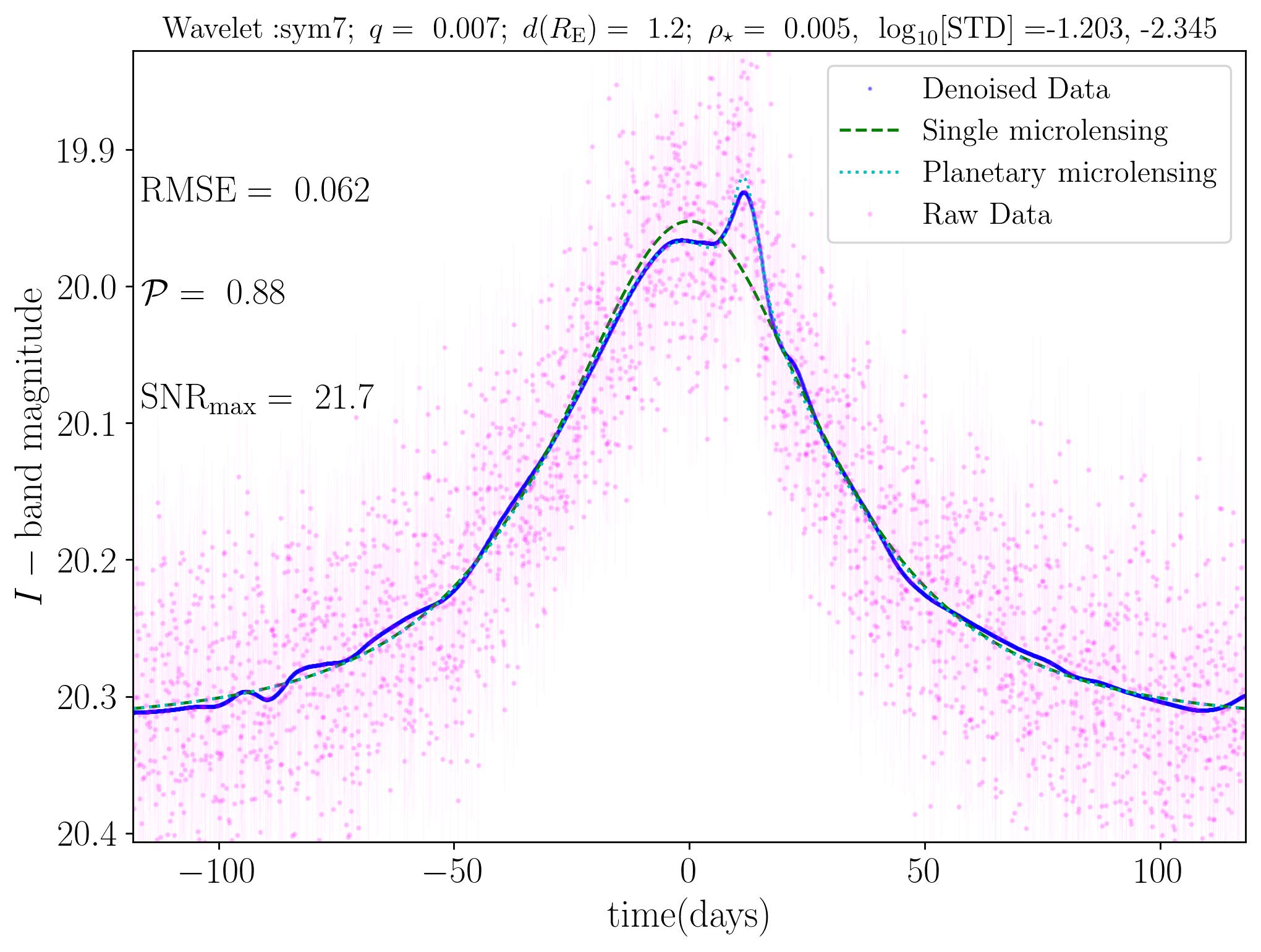}
\includegraphics[width=0.49\textwidth]{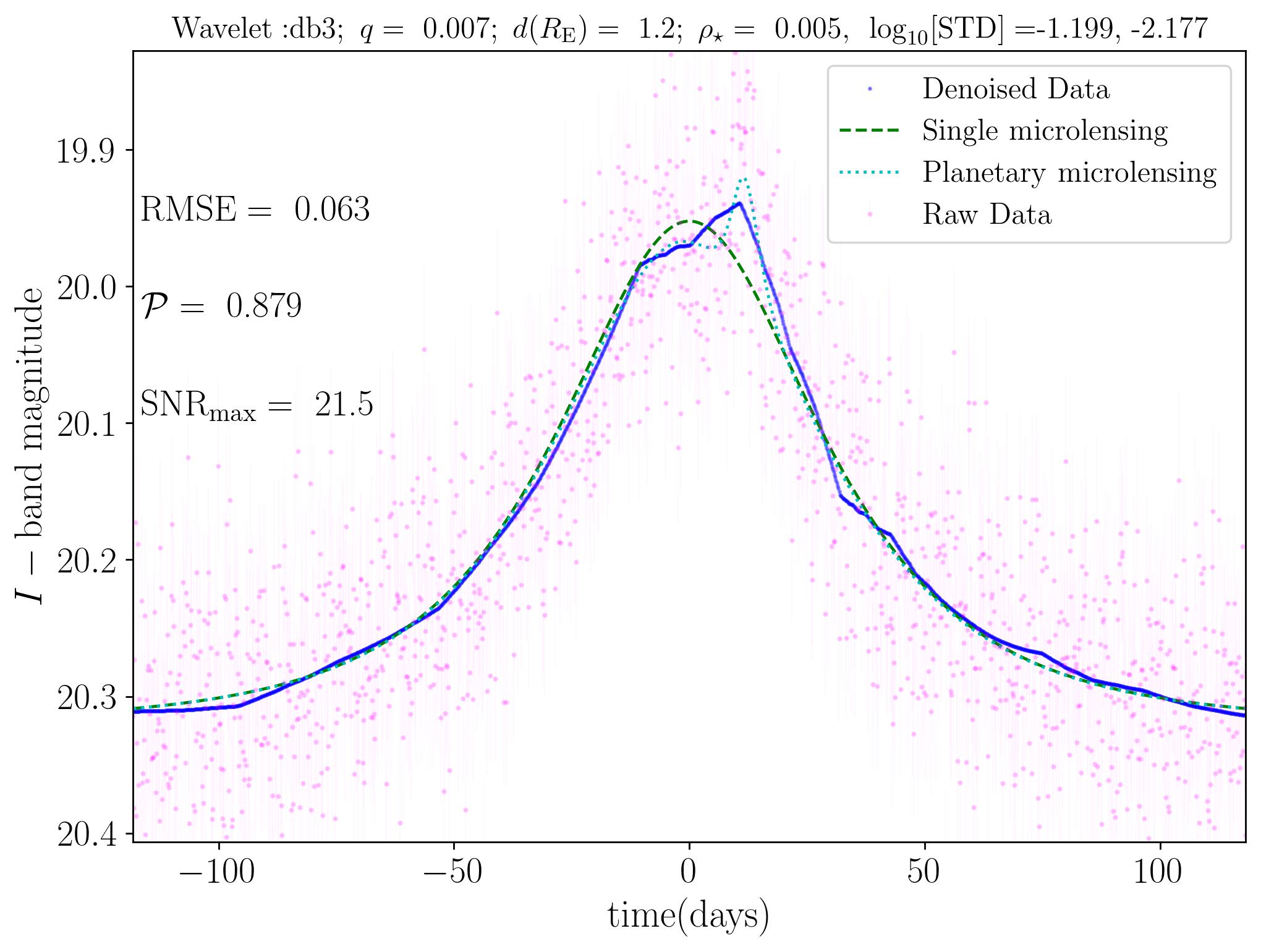}
\caption{Same as Figure \ref{Figone}, but with four different cadence values: $15$min (top left panel),  $1$ hour (top right panel), $2$ hours (bottom left panel), and $4$ hours (bottom right panel).}\label{Figtwo}
\end{figure*}
\begin{figure}
\centering
\includegraphics[width=0.49\textwidth]{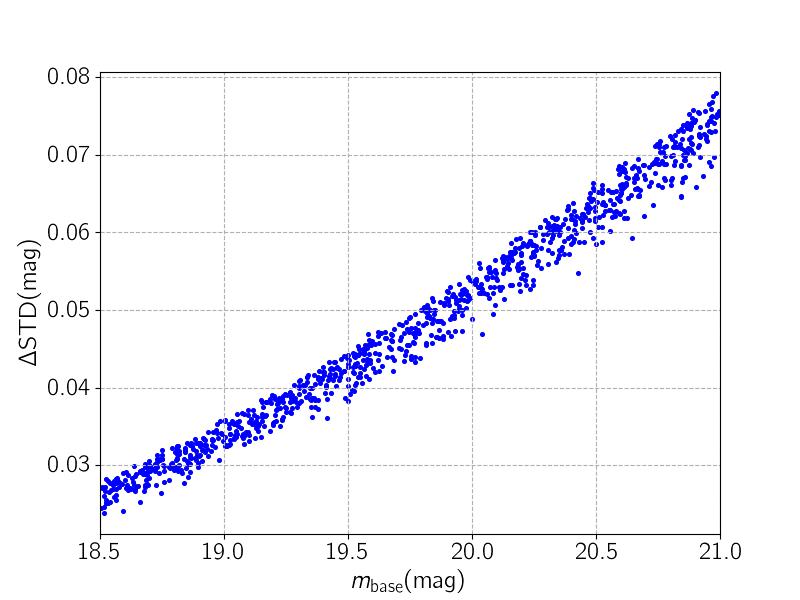}
\caption{The scatter plot of $\Delta \rm{STD}(\rm{mag})$ (improvements in STD of data due to DWT denoising) versus the baseline magnitude of the source stars for simulated planetary and single microlensing events.}\label{Dmbase}
\end{figure} 

The last criterion is the peak of SNR which is given by:  
\begin{eqnarray}
\rm{SNR}_{\rm{max}}= 10~\log_{10}\Big(\frac{\rm{Signal}_{\rm{max}}^{2}}{\rm{MSE}}\Big),
\label{cri3} 
\end{eqnarray}
where, MSE is the Mean Squared Error over data, and $\rm{Signal}_{\rm{max}}$ is the maximum possible value for $y_{i}$s. The larger $\rm{SNR}_{\rm{max}}$, the more reliable denoised data set.   
\begin{figure*}
\centering
\includegraphics[width=0.49\textwidth]{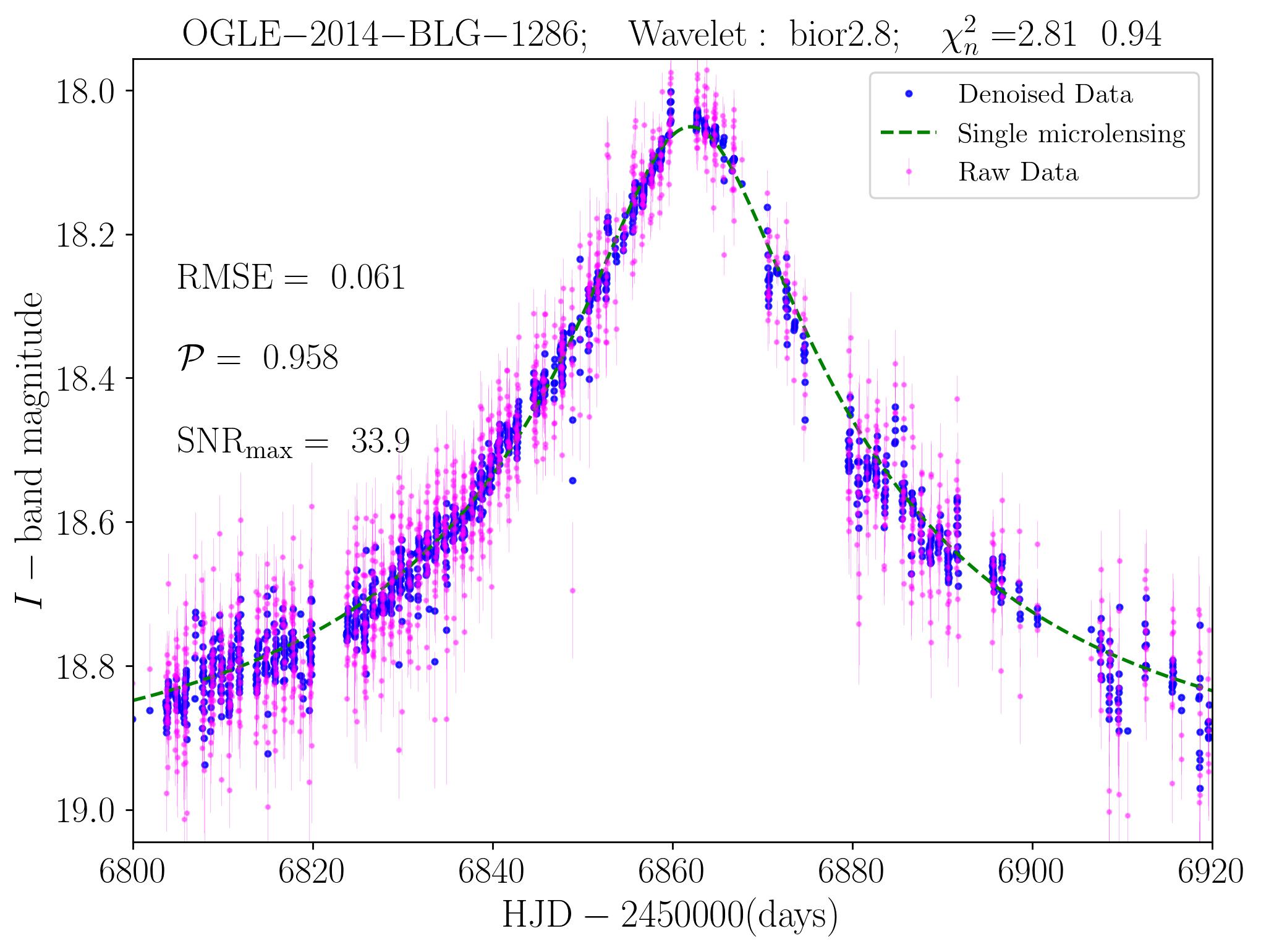}
\includegraphics[width=0.49\textwidth]{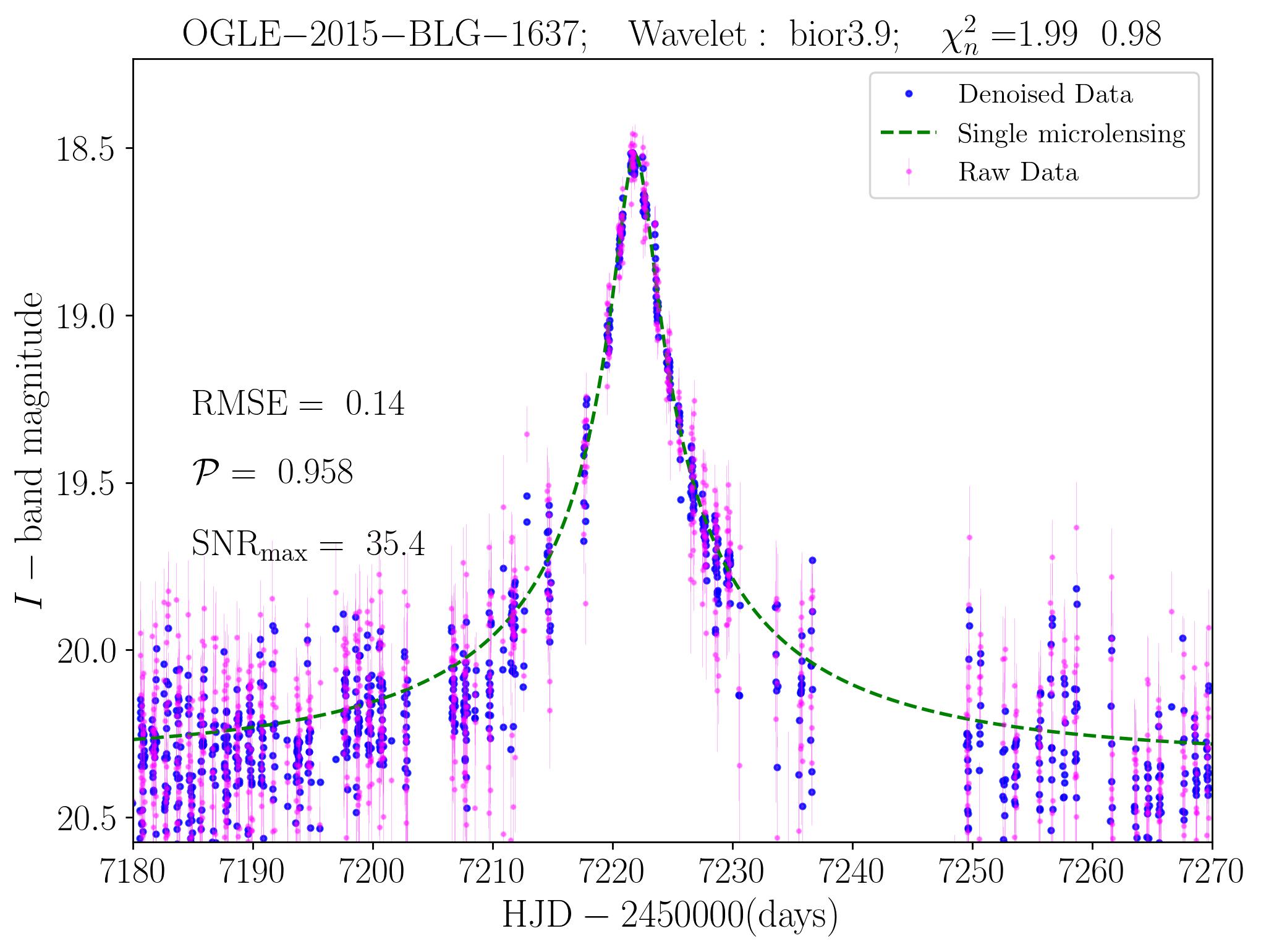}
\includegraphics[width=0.49\textwidth]{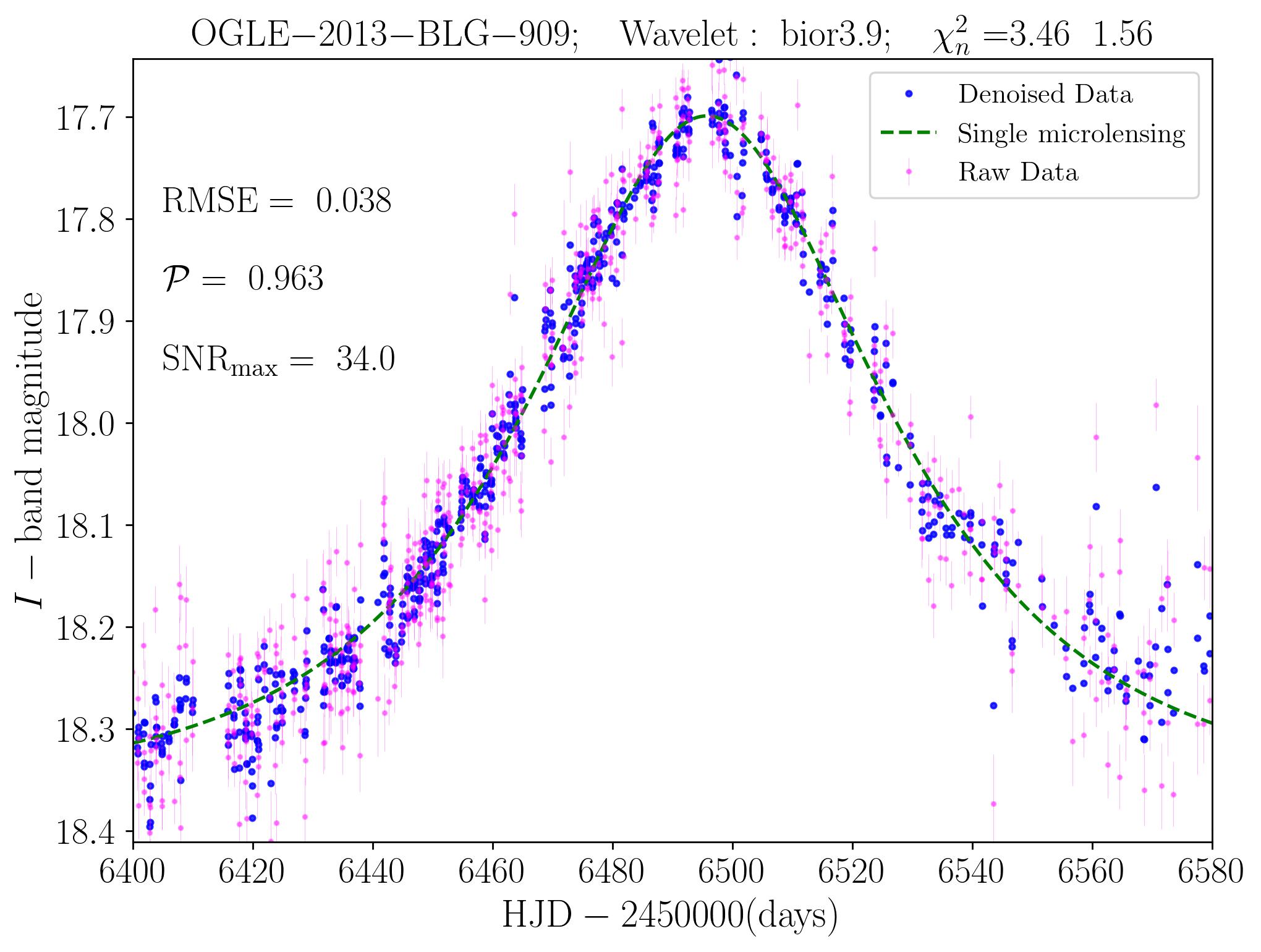}
\includegraphics[width=0.49\textwidth]{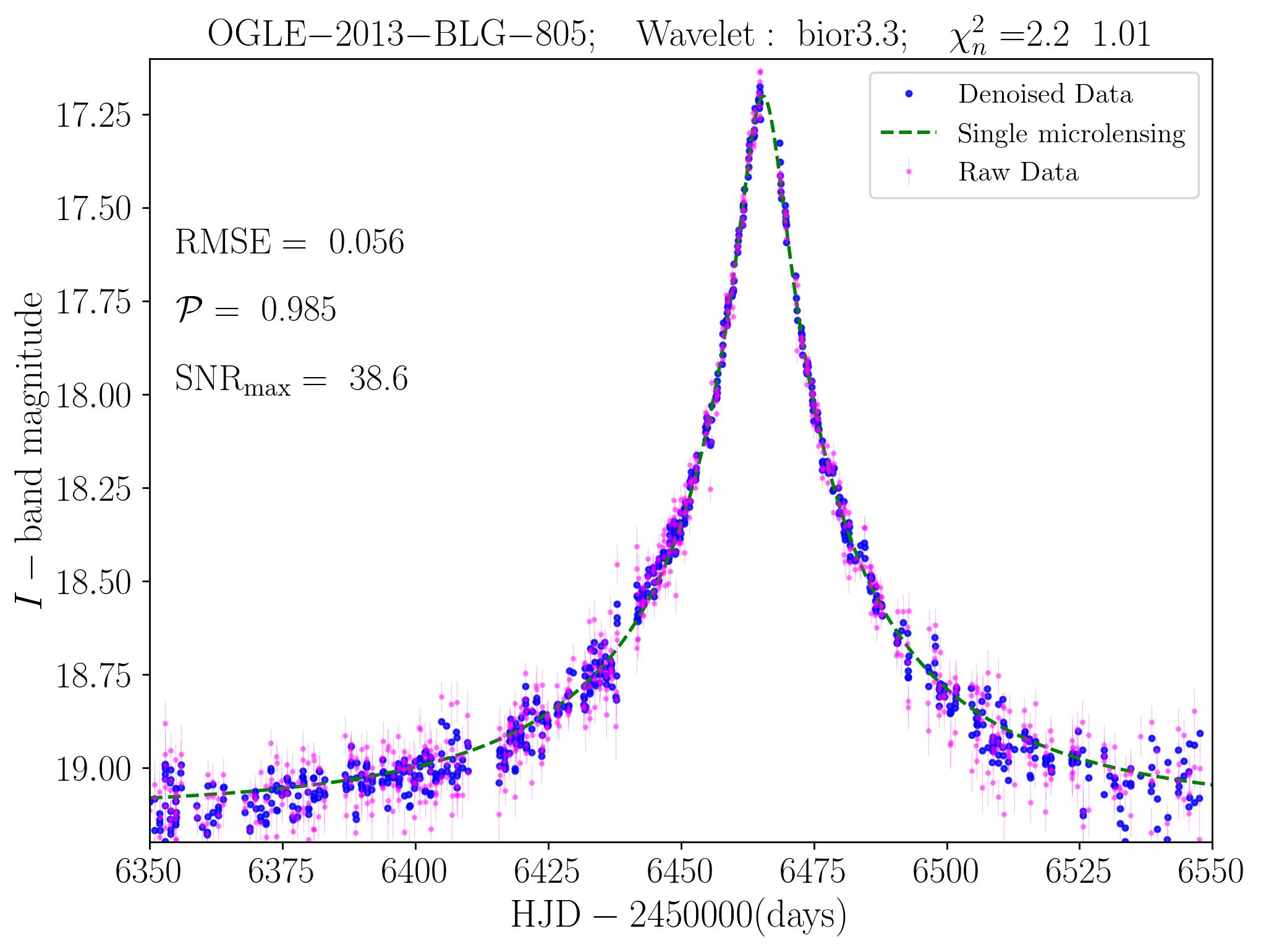}
\caption{Examples of denoised data due to four real microlensing events found by the OGLE group. The OGLE name of each event, the best-performing wavelet, and the amounts of $\chi^{2}_{\rm n}$ for raw, and denoised data (respectively) are mentioned at the top of each plot. The best-fitted single microlensing light curves (as characterized in the OGLE page) are shown with dashed green curves.}\label{Fig3}
\end{figure*}
\begin{figure}
\centering
\includegraphics[width=0.49\textwidth]{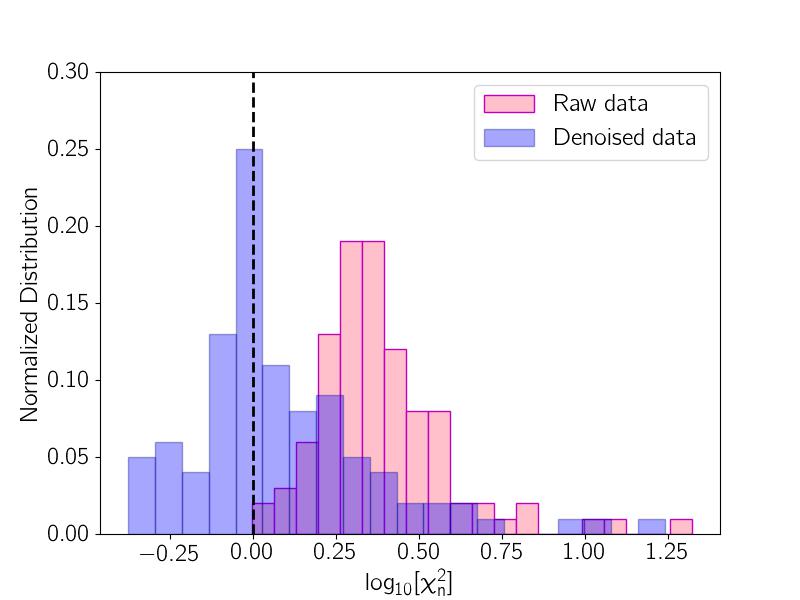}
\caption{The normalized distributions of $\chi^{2}_{\rm n}$s for raw (pink one) and denoised (blue one) data due to $100$ OGLE microlensing events.  }\label{Figchi}
\end{figure}

For each simulated microlensing event we visually inspect three denoised data sets due to wavelets with the minimum RMSE, the maximum $\mathcal{P}$, and the maximum $\rm{SNR}_{\rm{max}}$, and choose the best one. However, from simulations we find that these three criteria usually choose similar wavelets, and offer denoised data sets similar to raw data.

\noindent We note that it is not \textit{always} a good idea to choose the denoised data that very closely resembles the original data. Sometimes and when the source star is bright, the most similar denoised data set is exactly the original data. Therefore, we visually inspect the denoised data using all wavelets, and all thresholding methods, specially when a signal appears in denoised data set. In these cases, we inspect other denoised data sets if that signal exists in others.  

For simulated events we have their real models, so STD of denoised data from the real model is a reliable criterion to evaluate the improvement by wavelet denoising. We calculate this criterion by: 
\begin{eqnarray}
\rm{STD}(\rm{mag})= \sqrt{\frac{1}{N-1}\sum\limits_{i=1}^{N}  \Big(y_{\rm d,~i}-y_{\rm m, ~i}\Big)^{2} }, 
\label{std}
\end{eqnarray}  
where $y_{\rm m,~i}$s are the source apparent magnitudes in real models. This parameter shows how much denoised data describes the real model. 

In Figure \ref{Figone}, two examples of simulated planetary microlensing events with raw (pink ones), and denoised data (blue ones) are represented. In each panel, the real planetary microlensing model and the corresponding single microlensing one (obtained by fixing $q=0$) are shown with dotted cyan, and dashed green curves, respectively. The best-performing wavelets for denoising (based on the predetermined criteria), and the used planetary parameters are mentioned at the top of panels. Also, two given values for $\log_{10}\big[\rm{STD}(\rm{mag})\big]$ are due to raw and denoised data, respectively. In these two events, we generate synthetic data points with a $15$-min cadence.  
In both of these events, planetary signals are flattened and created small asymmetric features in light curves. Denoised data discerned the real models (depicted with dotted cyan curves) and reconstructed these signals well.

For $1000$ simulated planetary microlensing events, we extract the best denoised data sets according to the mentioned criteria and by visually inspecting. In Table \ref{tab1} the results are reported. In simulations, we found that the cadence is an important factor in the wavelet denoising and remaking real models. Hence, in the simulation we examine several values for observing cadence which are $15$ min, $1$ hr, $2$ hrs, $4$ hrs, and $6$ hrs as specified in different rows of Table \ref{tab1}. In this table, the second column $\epsilon$ determines the fraction of events in which their denoised data have $\log_{10}\big[\rm{STD}(\rm{mag})\big]<-2.79$ to the total number of simulated events in percent. We determined the threshold $-2.79$ by visually inspecting denoised data. The denoised data sets with $\log_{10}\big[\rm{STD}(\rm{mag})\big]<-2.79$ regenerate planetary signals well (see Figure \ref{Figone}).

\noindent The third column is $\overline{\Delta \rm{STD}}(\rm{mag})$, i.e., the average improvement in STD values (Eq. \ref{std}) due to denoising. The average is done over all simulated events. Two last columns specify the most-frequent wavelets and thresholding methods for the best denoised data sets.         

Accordingly, in continuous observations improving the observing cadence from $6$ hrs to $15$ minutes (i) enhances the efficiency to regenerate real models by denoising from $0.01\%$ to $37\%$, and (ii) decreases scattering of data with respect to real models by $0.004$ mag. As an example, in Figure \ref{Figtwo} we show a simulated planetary event with synthetic data points, and by considering four cadence values: $15$ minutes (top left panel), $1$ hr (top right panel), $2$ hrs (bottom left panel), and $4$ hrs (bottom right panel). In this event, there is a planetary signal which is fully recovered by denoised data with a $15$-min cadence. This planetary signal in denoised data with a $1$-hr cadence has a smaller peak. For two worse cadences, denoised data do not fully regenerate the planetary signal and the light curve's domains. 

The improvement in the STD of data due to denoising process strongly depends on the baseline magnitude of the source stars, as shown in Figure \ref{Dmbase}. According to this plot, DWT denoising improves the scattering of data for microlensing events from faint source stars rather than events from bright stars. We note that other parameters (e.g., the lens impact parameter, the Einstein crossing time) have smaller correlations with $\Delta\rm{STD}(\rm{mag})$.

By visually inspecting all denoised simulated events, we conclude that sharp and spark-like planetary signals are sometimes removed in the denoised data specially when data points are sparse, whereas flattened deviations are mostly distinguishable in denoised data. We note that in the simulation noises have an ideally Gaussian distribution. For these noises most of times the best-performing wavelets are from the 'Symlet' family.
  
In the next section, we apply denoising process on real observational microlensing data.  
\begin{figure*}
\centering
\includegraphics[width=0.49\textwidth]{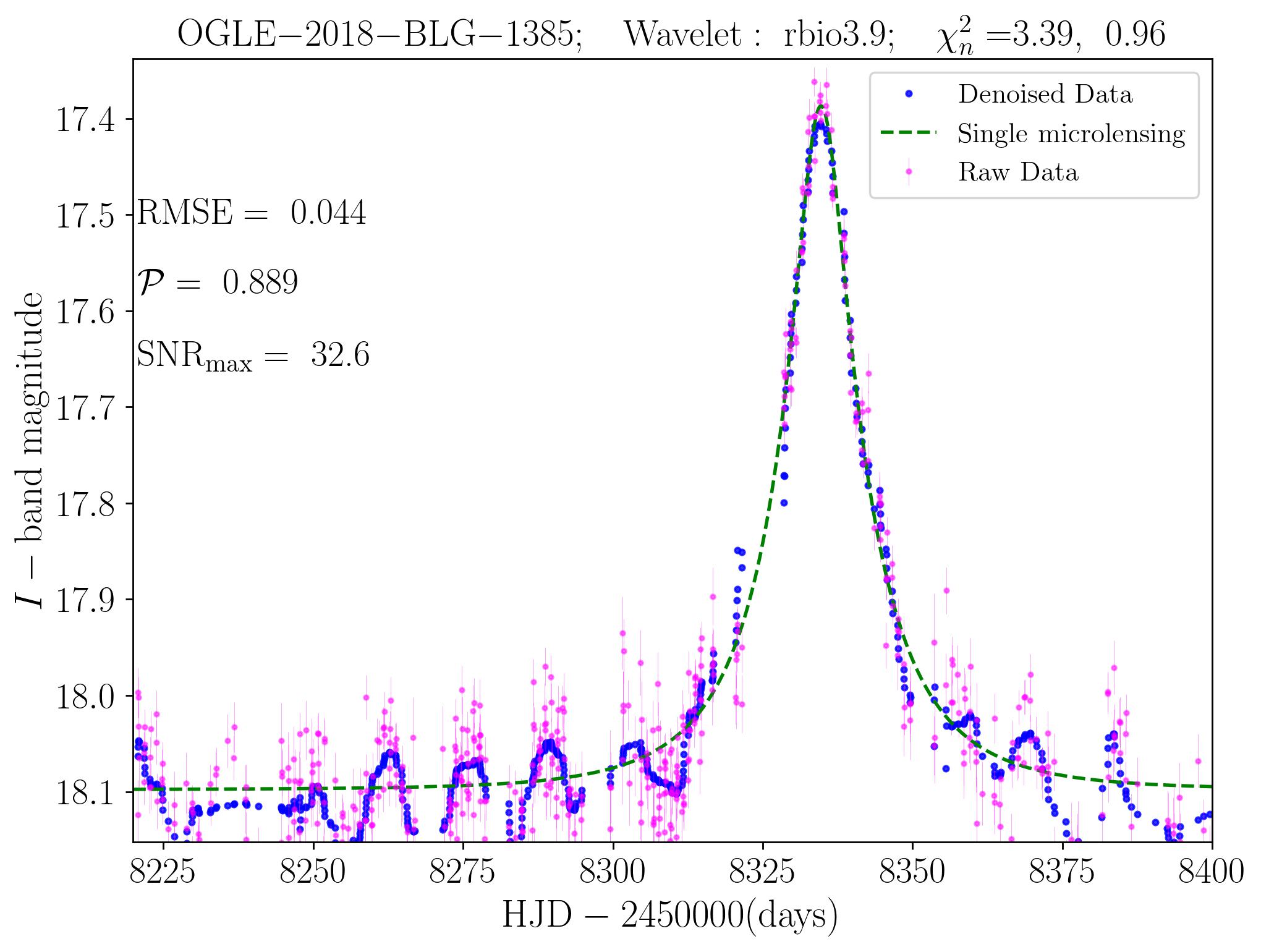}
\includegraphics[width=0.49\textwidth]{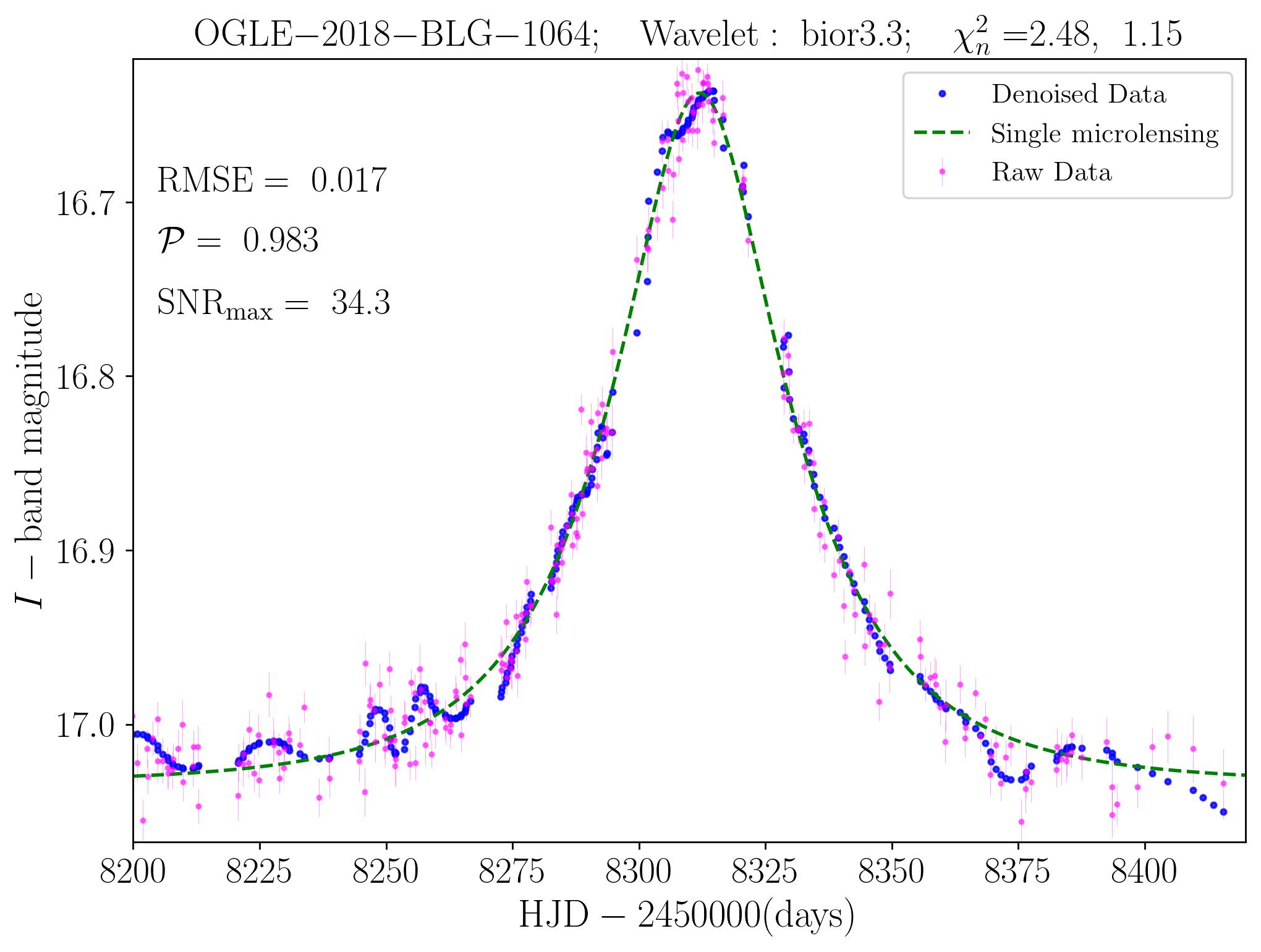}
\caption{Similar to Figure \ref{Fig3}, but in the denoised data of these two events intrinsic pulsations of source stars appear. These pulsations are not discernible in the raw data.}\label{Figpul}
\end{figure*}
\begin{figure*}
\centering
\includegraphics[width=0.49\textwidth]{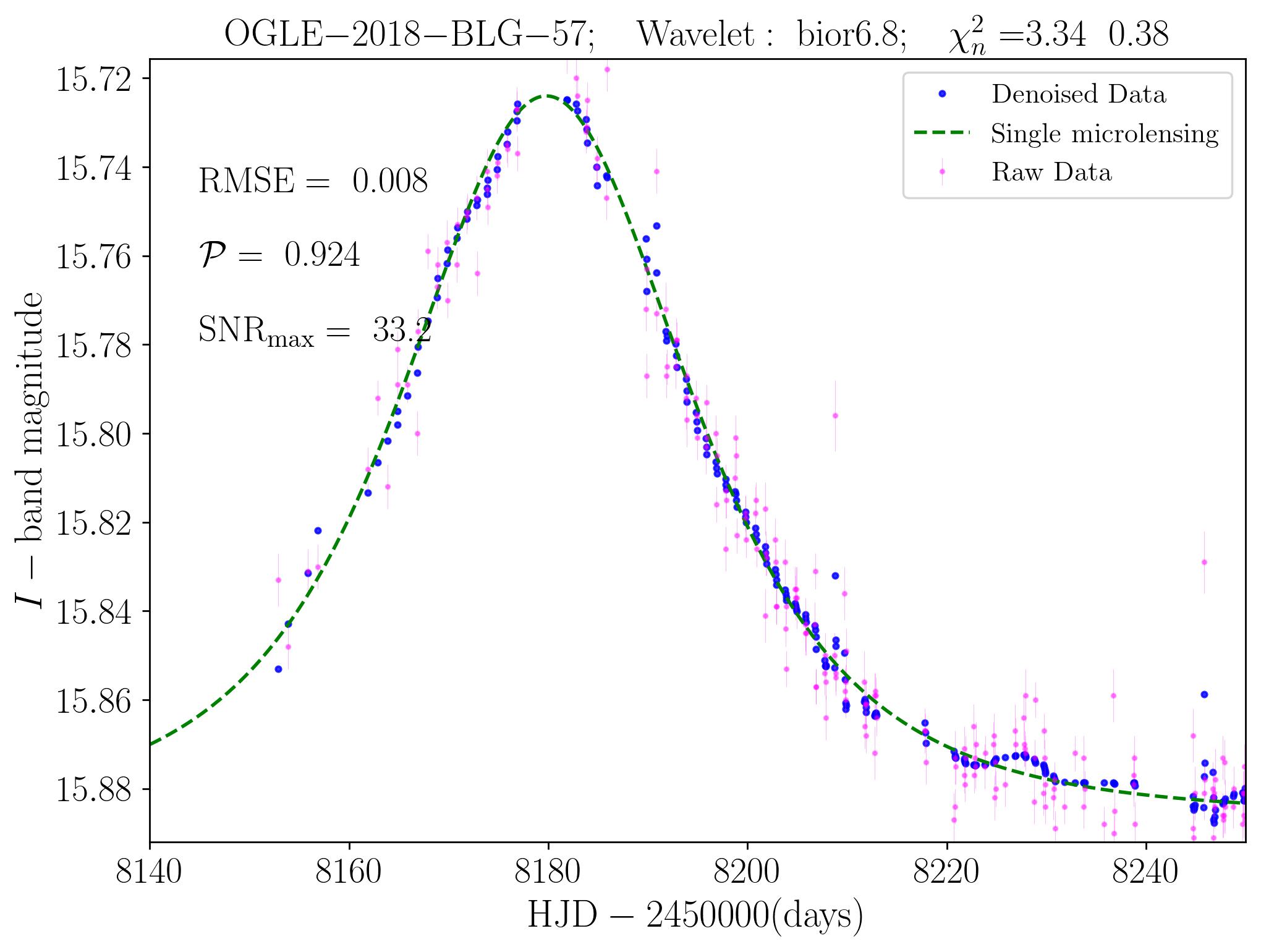}
\includegraphics[width=0.49\textwidth]{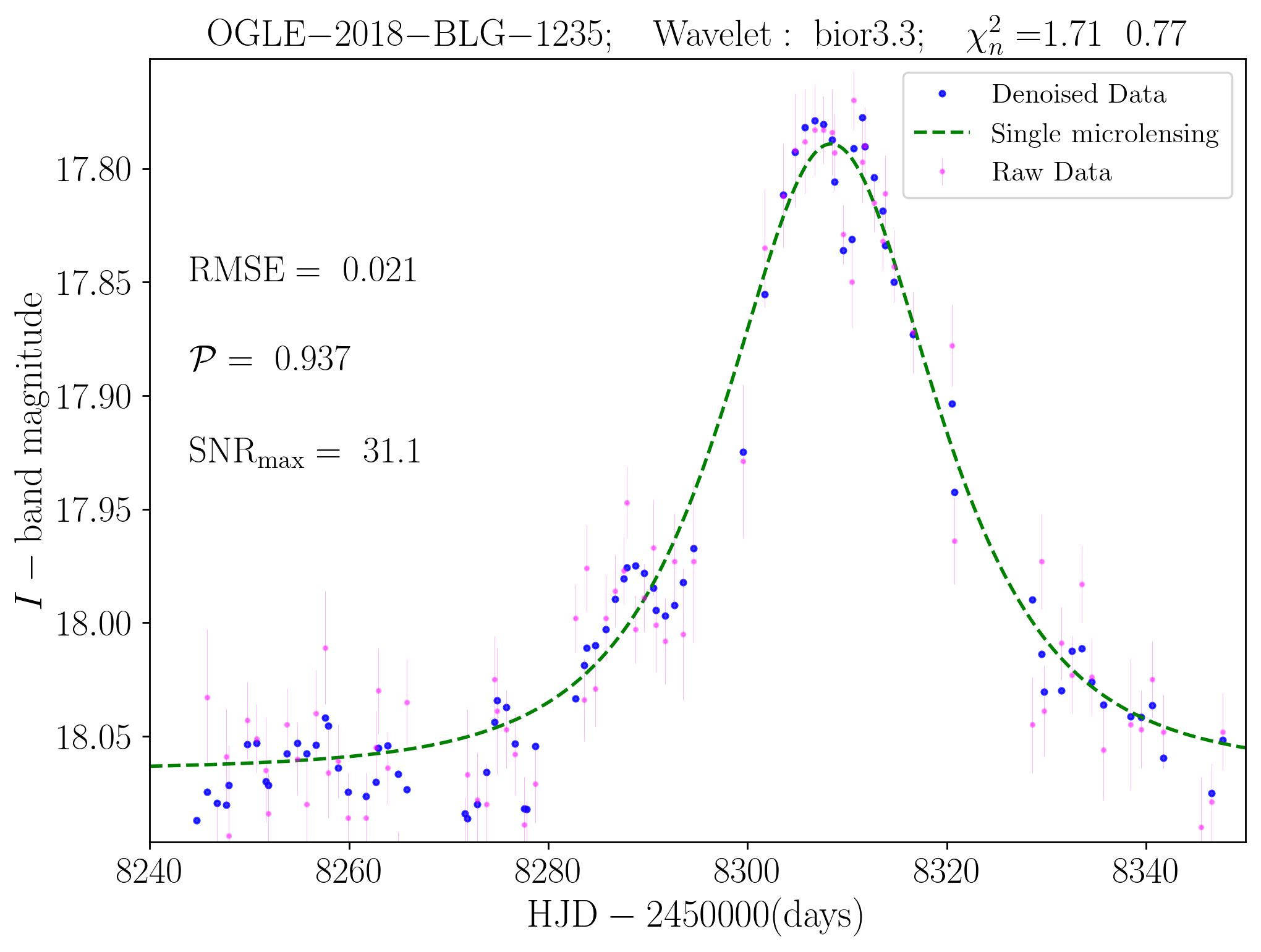}
\caption{Similar to Figure \ref{Fig3}, but for these two events, planetary-like deviations appear in the denoised data (blue ones), whereas these deviations are not discernible in raw data (pink ones). The planetary-like deviations are located at times $8225$, $8290$ days in the left and right-hand panels, respectively.}\label{Figplan}
\end{figure*}

\subsection{Real microlensing data}\label{real}
In this subsection, we apply a similar denoising process to $100$ microlensing events\footnote{Their list can be found here:  \url{https://iutbox.iut.ac.ir/index.php/s/epzM5mx67EMjyWX}} discovered by the OGLE group \citep{OGLE_IV,OGLE2003_1}. To evaluate the improvement by denoising process for these events, we need a criterion. For these events we do not know the real models, and so assume the proposed single microlensing models by the OGLE Early Warning System\footnote{\url{https://ogle.astrouw.edu.pl/ogle4/ews/ews.html}} are the best ones describing data. For an ideal data set, the value of $\chi^{2}$ of data with respect to the best-fitted model normalized to the number of data points should be one. Hence, for raw and denoised data sets we calculate this factor which is:
$$\chi^{2}_{\rm n}=\frac{1}{N}\sum\limits_{i=1}^{N} \Big(\frac{y_{i}-y_{\rm m,~i}}{\sigma_{i}}\Big)^{2},$$ 
where $\sigma_{i}$s are the photometric errors of data. Denoising process will improve $\chi^{2}_{\rm n}$. For the events without any significant deviations from a single microlensing light curve, by denoising process $\chi^{2}_{\rm n}$ gets closer to one.

In Figure \ref{Fig3}, we represent four examples of denoised (blue), and raw (pink) data due to microlensing events discovered by the OGLE group. In each panel, the best-fitted single microlensing light curve (characterized in the OGLE Early Warning System) is plotted by dashed green curves. At top of each panel, the OGLE name of that event, the best-performing wavelet, and $\chi^{2}_{\rm n}$ due to raw, and denoised data (respectively) are mentioned. In all of these events, denoising process makes data get closer to the models, which offers $\chi^{2}_{\rm n}$ to get closer to one.

We note that choosing the best-performing wavelet for each event is done based on three mentioned criteria (given by Equations \ref{cri1}, \ref{cri2}, and \ref{cri3}). By denoising these $100$ microlensing events, we find that $\chi^{2}_{\rm n}$ for the best denoising data sets on average decreases from $2.85$ to $1.69$. In Figure \ref{Figchi}, we show the normalized distributions of $\chi^{2}_{\rm n}$ values due to raw (pink distribution) and the best-denoised (blue one) data sets from these $100$ OGLE microlensing events. In this plot, the dashed black line specifies the ideal value of $\chi^{2}_{\rm n}=1$. The improvement in $\chi^{2}_{\rm n}$ values because of the denoising process is obvious from this plot. Also, the average STD of the OGLE microlensing data from the best-fitted models reduces from $0.074$ mag to $0.051$ mag because of denoising.

According to the results from Subsection \ref{simul}, for the simulated microlensing data with ideally Gaussian noises the best-performing wavelets are from 'Symlet' wavelet family most of time. On the other hand, for real microlensing data the best-performing wavelets are from 'Biorthogonal' wavelet family. It means that observing noises do not obey from an ideally Gaussian distribution. Also we note that real observations are not continuous. 

\noindent We also note that Symlet wavelets are a modified version of Daubechies wavelets with increased symmetry which are orthogonal, whereas Biorthogonal wavelets (made from nonorthogonal bases and dual terms) have more flexibility and less symmetry. Simulated data with ideally Gaussian noises are more symmetric than real data. Because in real data usually there are some discontinuities and some locally high-noisy characteristics. Therefore, for real data Biorthogonal wavelets will model noises better than Symlet wavelets.

One advantage of wavelet denoising of microlensing data is that intrinsic stellar pulsations are potentially emerged in the denoised data. Two examples of such events are depicted in Figure \ref{Figpul}. In both events, stellar pulsations are not recognizable in the raw data, whereas by denoising these pulsations appear. We note that discerning stellar pulsations could be well done for continuous data without large gaps (see, the right-hand panel in Fig. \ref{Figpul}). Intrinsic pulsations change the shape of light curves, and as a results, modeling of microlensing events with noisy stellar pulsations will offer wrong best-fitted parameters \citep[see, e.g., ][]{sajadian2020a,2020sajadianbb,2021sajadian}.

For two OGLE microlensing events in our sample, wavelet denoising makes two planetary-like signals appear in the denoised data, whereas these signatures were not discernible in their raw data. In Figure \ref{Figplan}, we show these two events. The planetary-like signals happen at times $8225$, and $8290$ days for left and right-hand light curves, respectively. We note that these deviations appear not only in the denoised data set with the best-performing wavelet, but also in other denoised data sets (with other wavelets). For that reason, these signatures are trustful. These deviations show the importance of denoising process for microlensing data. Therefore, weak planetary-like signatures in noisy microlensing data could be extracted through wavelet denoising process.      
\begin{figure*}
\centering
\includegraphics[width=0.49\textwidth]{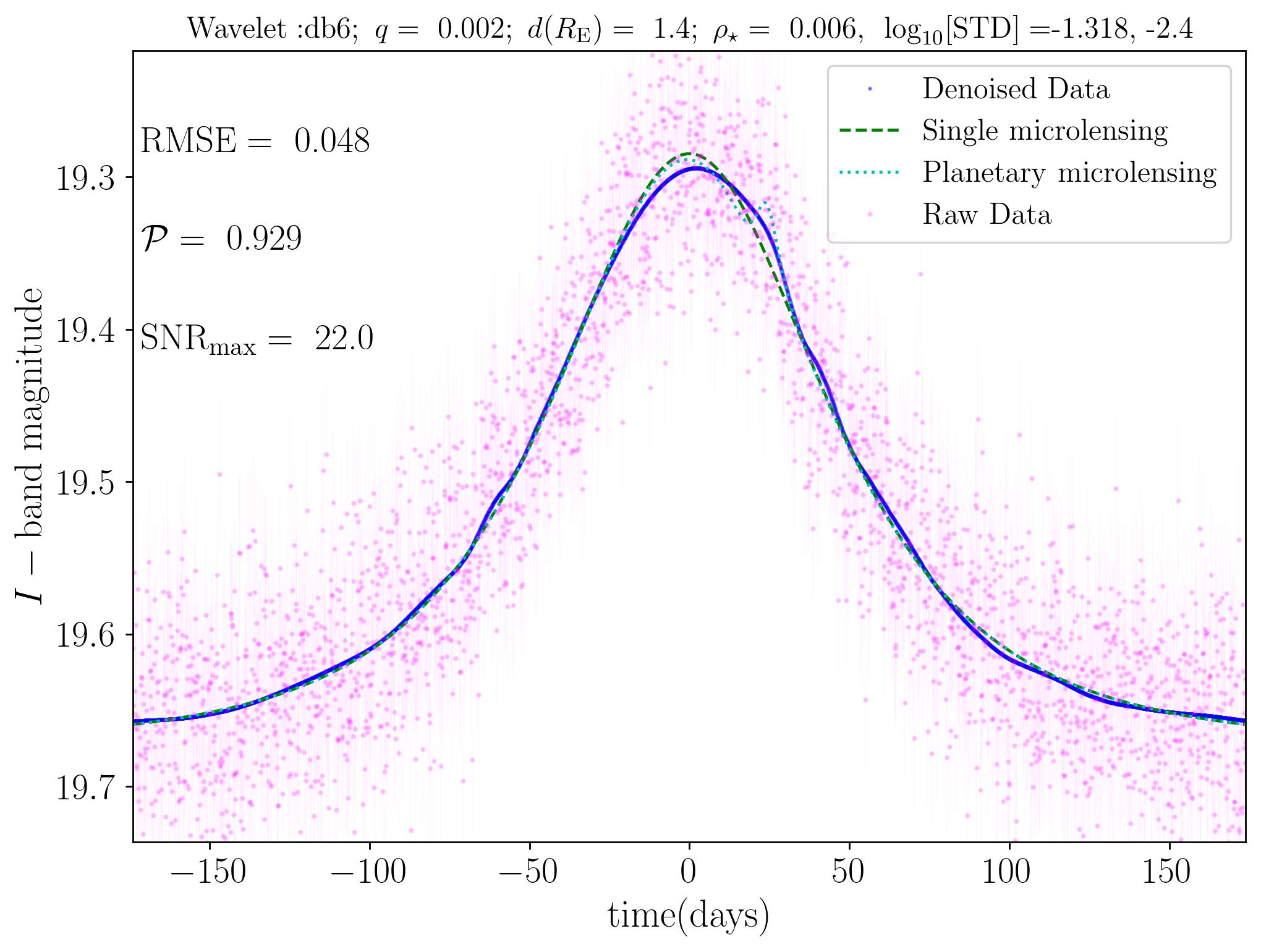}
\includegraphics[width=0.49\textwidth]{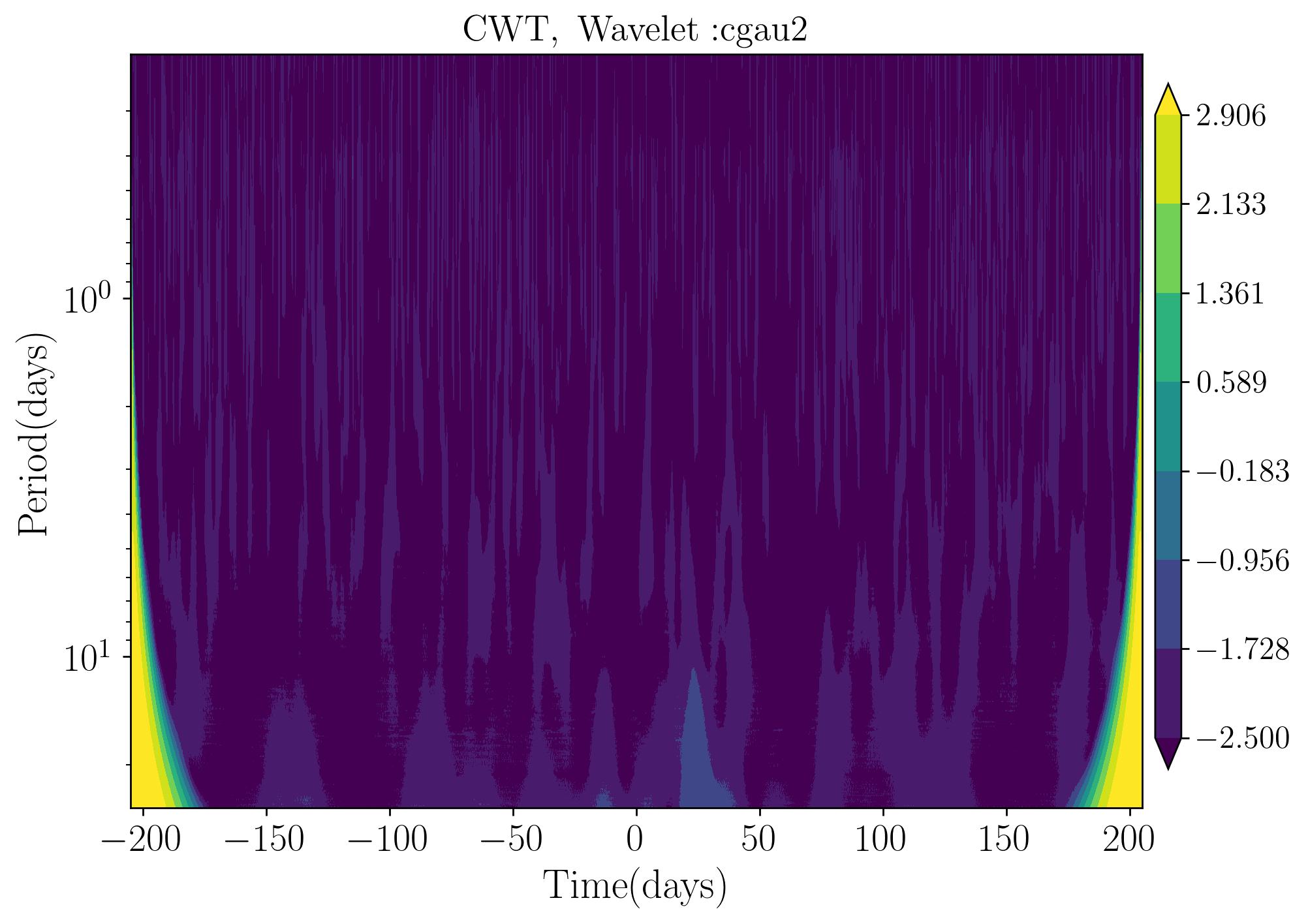}
\caption{Left panel: Similar to Figure \ref{Figone}, by considering a $2$-hour cadence for simulating data. Right panel: Its 3D wavelet map resulted from CWT with the wavelet 'cgau2'.}\label{Figlst}
\end{figure*} 
\begin{figure*}
\centering
\includegraphics[width=0.49\textwidth]{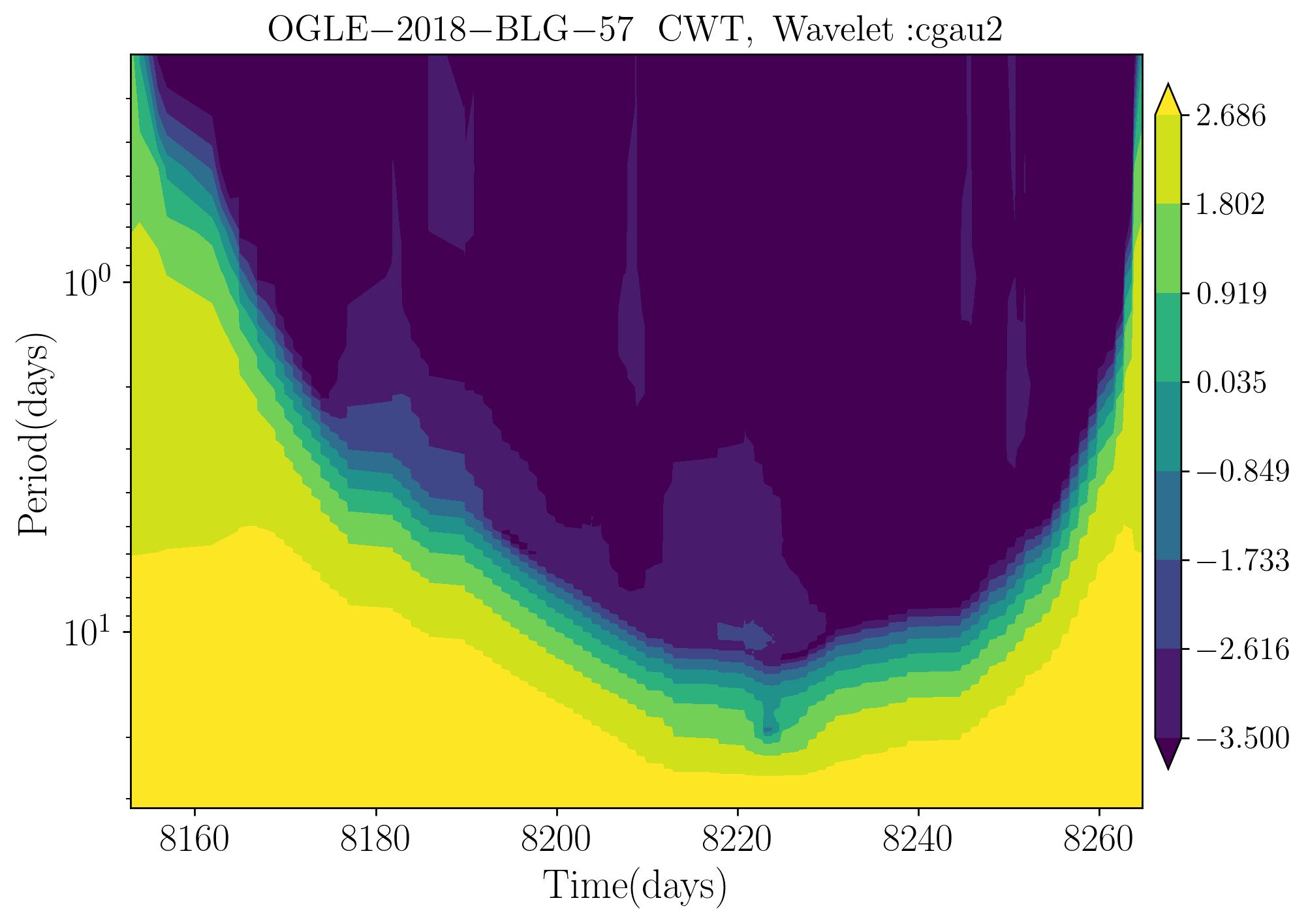}
\includegraphics[width=0.49\textwidth]{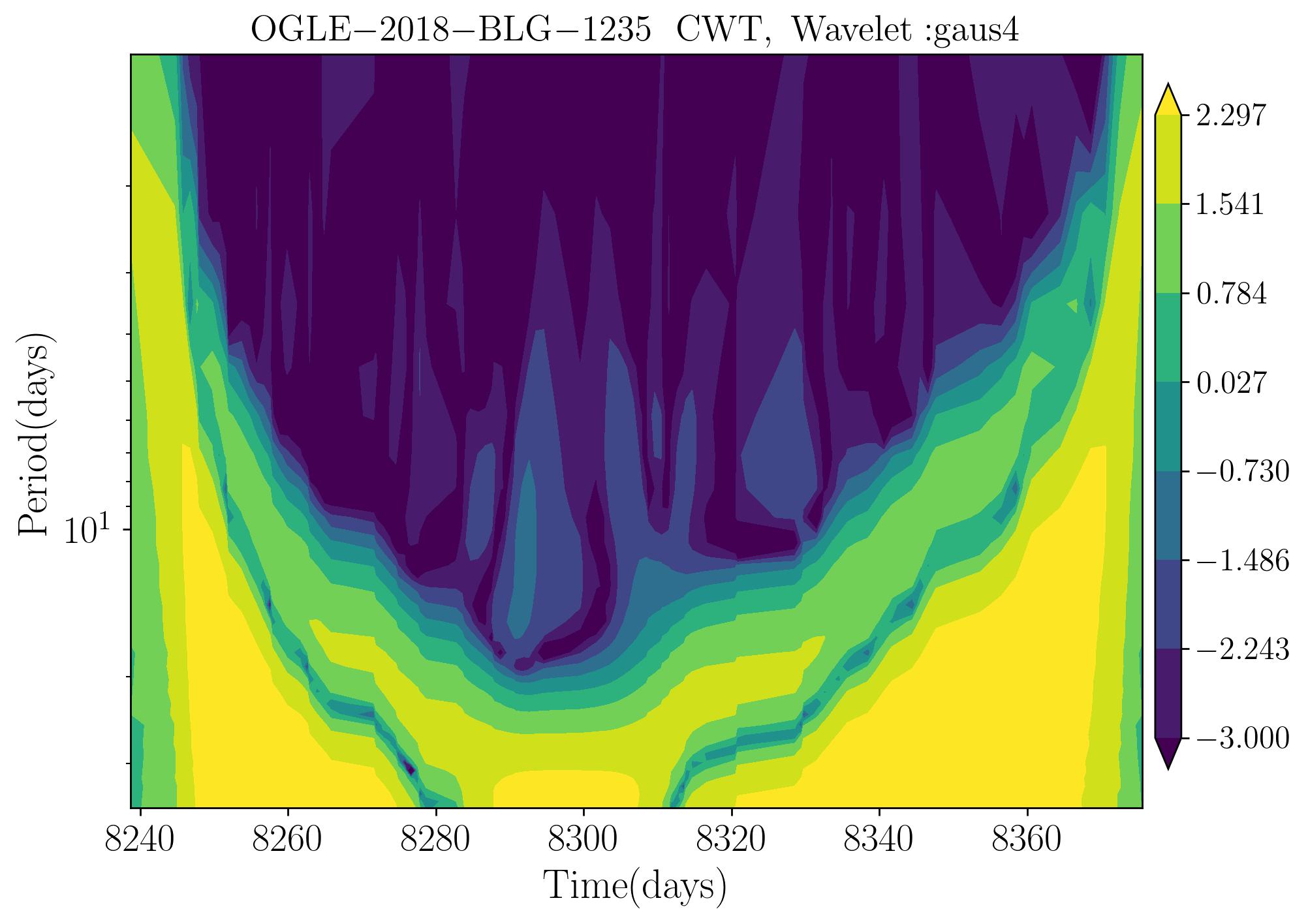}
\caption{3D wavelet maps resulted from CWT for two OGLE events shown in Figure \ref{Figplan}, with emerged planetary-like signatures in their denoised data.}\label{Figmap}
\end{figure*} 

\noindent In Subsection \ref{simul}, we found this point by applying denoising process on simulated planetary microlensing events (two examples can be found in Figure \ref{Figone}). Also we noticed that denoising sometimes (specially when the observing cadence is long) removes spark-like and sharp planetary deviations with short durations, although it discerns wide and flattened planetary signals well such as ones shown in Figure \ref{Figplan}. In the next section, we express these sharp planetary deviations in microlensing light curves could be distinguished in the wavelet power spectrum resulted from the continuous wavelet transform.    

\section{CWT of microlensing data}\label{cwt}
Continuous Wavelet Transform (CWT) is a very helpful tools to discern temporary and unstable deviations in time series \citep[e.g., ][]{2014aabravo}. Using CWT, we can plot 3D frequency-power-time map (the wavelet map). For simulated microlensing events, we apply CWT on raw data using the python module \texttt{PyWavelet}, and plot their wavelet maps. The continuous wavelet families are 'Gaussian (gaus)', 'Mexican hat(mexh)', 'Morlet(morl)', 'Complex Gaussian(cgaus)', 'Shannon(shan)', and 'Complex Morlet(cmor)', used for CWT.

For a single microlensing event without any asymmetric deviations, its wavelet map is symmetric with respect to the time of the closest approach. For planetary events with asymmetric deviations, the resulted wavelet map is not symmetric. One example of such events is shown in the left panel of Figure \ref{Figlst}. In this simulated microlensing event, there is a small planetary deviation at the time $25$ days (see the dotted cyan curve), which was not reconstructed by denoised data (blue points). Its wavelet map using CWT is shown in the right panel of Figure \ref{Figlst}. In this map the color bar shows the power of each wavelet basis in the wavelet decomposition form. We calculate powers as $|C_{a,~b}|^{2}$. The used continuous wavelet is mentioned at the top of the panel. The wavelet map is symmetric with respect to $t_{0}$, unless there are some weak features around the peak and specially at $t=25$ days (at the position of the planetary-induced deviation). Hence, CWT map potentially informs us about the existence of planetary deviations. However, it can not remake the shape of these deviations.

To understand whether plotting 3D CWT maps is in reality helpful for real microlensing observations or not, in Figure \ref{Figmap} we plot the wavelet maps of two OGLE microlensing events shown in \ref{Figplan}. For the real microlensing events, the wavelet maps may be asymmetric because of discontinuity in their data, but not due to planetary signals. According to wavelet maps in Figure \ref{Figmap},  there are some features at times $t_{0}$ (i.e., $8180$, and $8310$ days which specify the magnification peaks), $8220$ (left panel), and $8290$ days (right panel). Hence, by plotting 3D wavelet maps from CWT of time-series microlensing data one can validate weak signals, as well.  

\small

The list of OGLE events studied in Section \ref{real}, simulations, and codes have been developed for this work can be found in these GitHub and Zenodo addresses:  \url{https://github.com/SSajadian54/DE-NOISE}, and \url{https://zenodo.org/record/8403741}\citep{SajadianFatheddin2023}.

\section{Conclusions}\label{conclusion}
Wavelets are waveform and localized mathematical functions with finite energy, and zero mean. Hence, these functions well model transient and unstable noises in time-series data. By expanding a given function versus wavelet bases, one can denoise that function and extract its signals \citep[e.g. see ][]{Morlet1982a,Morlet1982b}. Denoising time-series data and extracting their signals using wavelet transforms have been used frequently in various fields of astronomy, astrophysics, and even cosmology \citep[For example,][]{2003GRBwavelet,cmbwavelet,2014aabravo}. As a new wavelet application, in this work we studied advantages of DWT and CWT of simulated and real microlensing time-series data.  

We first studied DWT denoising of microlensing data. We simulated $1000$ planetary and single microlensing events by considering the OGLE photometric accuracy, and applied DWT denoising process to them. For each event, we chose the best-performing wavelet and the best thresholding method based on three statistical criteria, i.e., RMSE, Pearson's correlation, and maximum of the SNR's peak (given by Equations \ref{cri1}, \ref{cri2}, and \ref{cri3}, respectively). The improvement due to DWT denoising was evaluated by the STD values of denoised data sets with respect to their real models (given by Equation \ref{std}). The average improvement in STD values because of denoising is $0.044$-$0.048$ mag.  

\noindent In the simulation, we noticed that in the events with $\log_{10}[\rm{STD}(\rm{mag})]<-2.79$ the denoised data would predict real models and reproduce planetary signals well. The fraction of these events to total simulated events, $\epsilon$, depends strongly on the cadence and decreases from $37\%$ to $0.01\%$ by worsening cadence from $15$ minutes to $6$ hrs.% in continuous observations. 

We performed DWT denoising for $100$ single microlensing events discovered by the OGLE group. For these events we evaluated the denoising impacts using $\chi^{2}_{n}$ of data from the best-fitted models characterized in the OGLE web-page. DWT denoising improved $\chi^{2}_{\rm n}$ and STD of these OGLE microlensing data by on average $1.16$ and $0.023$ mag, respectively. The best-performing wavelets (based on three mentioned criteria) are from 'Symlet', and 'Biorthogonal' wavelets families in simulated, and OGLE microlensing data, respectively.

\noindent In some OGLE events, stellar intrinsic pulsations or planetary-like signals appeared in the denoised data which were covered with noises in raw data (see, Figure \ref{Figpul}, \ref{Figplan}).

Additionally, we found that DWT denoising of microlensing data could reconstruct rather flattened and wide planetary-like signals than sharp and spark-like ones. For discerning sharp planetary signals, CWT and generating 3D frequency-power-time maps potentially reveal their existences, although these maps cannot determine shapes of these planetary signals. 

\bibliographystyle{aasjournal}
\bibliography{paper}{}
\end{document}